\documentclass[prd,two column,nofootinbib,english]{revtex4-1}
\usepackage[T1]{fontenc}
\usepackage[latin1]{inputenc}
\usepackage{float}
\usepackage{babel}
\usepackage{graphicx}
\usepackage{setspace}
\usepackage{amsmath}
\usepackage{braket}
\usepackage{color}
\begin{document}
\def\sl#1{\slash{\hspace{-0.2 truecm}#1}}
\def\beqn{\begin{eqnarray}}
\def\eeqn{\end{eqnarray}}
\def\nn{\nonumber}
\def\bea{\begin{eqnarray}}
\def\eea{\end{eqnarray}}

%
%
\title{Beam normal spin asymmetry for the $e p \to e \Delta(1232)$ process}
\author{Carl E. Carlson}
\affiliation{College of William and Mary, Physics Department, Williamsburg, Virginia 23187, USA}
\author{Barbara Pasquini}
\affiliation{Dipartimento di Fisica, Universit\`a degli Studi di Pavia, 27100 Pavia, Italy}
\affiliation{INFN Sezione di Pavia, 27100 Pavia, Italy}
\author{Vladyslav Pauk}
\affiliation{Jefferson Laboratory, Newport News, VA 23606, USA}
\author{Marc Vanderhaeghen}
\affiliation{Institut f\"ur Kernphysik and PRISMA Cluster of Excellence, Johannes Gutenberg Universit\"at, D-55099 Mainz, Germany}
\date{\today}
\begin{abstract}
We calculate the single spin asymmetry for the $e p \to e \Delta(1232)$ process, for an electron beam polarized normal to the scattering plane. Such single spin asymmetries vanish in the one-photon exchange approximation, and are directly proportional to the absorptive part of a two-photon exchange amplitude. As the intermediate state in such two-photon exchange process is on its mass shell, the asymmetry allows one to access for the first time the on-shell $\Delta \to \Delta$ as well as $N^\ast \to \Delta$ electromagnetic transitions. We present the general formalism to describe the $e p \to e \Delta$ beam normal spin asymmetry, and provide a numerical estimate of its value using the nucleon, $\Delta(1232)$, $S_{11}(1535)$, and $D_{13}(1520)$ intermediate states.  
We compare our results with the first data from the Qweak@JLab experiment and give predictions for the A4@MAMI experiment.  
\end{abstract}
%
\maketitle 
\section{Introduction}

A lot of information is available on the electromagnetic structure of protons and neutrons,  
such as their magnetic moments, charge radii, elastic form factors, or electromagnetic polarizabilities. 
In contrast, our knowledge on the electromagnetic structure of nucleon excited states is very scarce. 
Even for the lowest excitation of the nucleon, the prominent $\Delta(1232)$ resonance, 
the information is limited to the non-diagonal $N \to \Delta$ electromagnetic transition, see 
e.g. Refs.~\cite{Pascalutsa:2006up,Alexandrou:2012da,Aznauryan:2012ba} for some recent reviews.  
Deducing from such measurements physical quantities as the magnetic dipole moment or 
the charge radius of the $\Delta(1232)$ state, has long required resorting to theoretical approaches which relate the 
properties of the  $\Delta$ to properties of the nucleon and/or to the experimentally accessible $N \to \Delta$ 
transition. Such theoretical approaches include different types of constituent quark models (see Ref.~\cite{Pascalutsa:2006up} for a review of some of these models), 
general large-$N_c$ relations in QCD~\cite{Jenkins:1994md,Lebed:2004fj,Buchmann:2000wf,Buchmann:2002et}, as well as 
chiral effective field theory including nucleon 
and $\Delta$ fields~\cite{Gellas:1998wx,Pascalutsa:2004je,Gail:2005gz,Procura:2008ze,Ledwig:2011cx}. 
In recent years, lattice QCD has been able to also provide direct calculations of such static quantities and  
FFs for the $\Delta$ resonance~\cite{Alexandrou:2008bn, Alexandrou:2009hs,Aubin:2008qp}.   

In order to experimentally access the electromagnetic structure of the $\Delta(1232)$ resonance,  
and to directly compare with lattice QCD predictions,  a way to measure the diagonal $\Delta \to \Delta$ electromagnetic transition is required. 
As the $\Delta(1232)$ is a very short lived resonance, the only viable way is to use a reaction where 
the $\Delta$ is first produced, and then 
couples to the electromagnetic field before decaying into a $\pi N$ state.  
One such process which has been proposed to access the magnetic dipole moment (MDM) of the $\Delta^+(1232)$ resonance 
is the radiative $\pi^0$ photoproduction process $\gamma p \to \gamma \Delta^+ \to \gamma \pi^0 p$ 
~\cite{Machavariani:1999fr, Drechsel:2000um, Drechsel:2001qu, Chiang:2004pw, Pascalutsa:2004je, Pascalutsa:2007wb}. 
A first experimental extraction of the $\Delta^+(1232)$ MDM has been performed in Ref.~\cite{Kotulla:2002cg} using the reaction model of Ref.~\cite{Drechsel:2001qu}, resulting in the value listed by PDG~\cite{Olive:2016xmw}: 
\begin{equation}
\mu_{\Delta^+} =  2.7 \mbox{${{+1.0} \atop {-1.3}}$}
(\mathrm{stat.}) \pm 1.5 (\mathrm{syst.}) \pm 3 (\mathrm{theor.})\, \mu_N\,,
\label{eq:mdm}
\end{equation}
with $\mu _N=e/2M_N$ the nuclear magneton. One notices that the error in Eq.~(\ref{eq:mdm}) is dominated by the theoretical 
uncertainty.  
A dedicated follow-up $\gamma p \to \gamma \pi^0 p$  experiment \cite{Schumann:2010js} found it difficult 
to improve on the precision of the $\Delta^+$ MDM due to model dependencies in the used theoretical framework, which is needed to access the on-shell $\gamma^\ast \Delta \Delta$ vertex from such reaction process.  

Accessing the on-shell electromagnetic FFs of the $\Delta(1232)$ 
resonance has not been possible in experiment to date.  
To achieve such goal, we need a two-photon observable where the 
$\Delta$ is firstly produced on a proton target by one virtual photon and then couples to the second photon leading to the $\Delta$ final state, which is then detected through its $\pi N$ decay. 
In order to properly access the on-shell $\gamma^\ast \Delta \Delta$ vertex, we need to look 
at  the pole-position  of the intermediate $\Delta$ state. If we want to realize such an experiment with virtual photons it will in general be dominated by the direct electromagnetic $N \to \Delta$ transition which involves only one photon, and is well studied in experiment, 
e.g. through the pion electroproduction process on a proton in the $\Delta$ region. 
If we aim to access the electromagnetic $\Delta$ FFs, we need an observable where this direct 
$N \to \Delta$ transition through one photon is suppressed or absent. 
An observable which realizes this is the 
beam normal spin asymmetry for the $e p \to e \Delta(1232)$ process, which we study in this work. 

Normal single spin asymmetries (SSA) for the $e p \to e R$ processes, with $R$ some well defined state, e.g. reconstructed through its invariant mass, with either the electron beam or the hadronic target 
polarized normal to the scattering plane, are exactly zero in absence of two or multi-photon exchange contributions. 
These normal SSAs are  proportional to the imaginary (absorptive) part of the 
two-photon exchange (TPE) amplitude, which is the reason why they are exactly zero for real (non-absorptive) processes such as one-photon exchange (OPE). 
At leading order in the 
fine-structure constant, $\alpha = e^2 / (4 \pi) \simeq 1/137$, the normal SSA results from the product between the OPE amplitude and the imaginary part of the TPE amplitude, 
see Ref.~\cite{Carlson:2007sp} for a recent review.  As the SSA is proportional to the imaginary part of the TPE amplitude at leading order in $\alpha$, it guarantees that the intermediate hadronic state is produced on its mass shell. 

For a target polarized normal to the scattering plane, the corresponding normal SSAs were predicted to be in the (sub) per-cent range some time ago~\cite{DeRujula:1972te}. Recently, a first measurement  of the target normal SSA for the elastic 
electron-$^3$He scattering  has been performed by the JLab Hall A Coll., 
extracting a SSA for the elastic electron-neutron subprocess, for a normal polarization of the neutron, in the per-cent range~\cite{Zhang:2015kna}.  
For the experiments with polarized beams, the corresponding normal SSAs for the $e p \to e p$ process  involve a lepton helicity flip which is suppressed by the mass of the electron relative to its energy. Therefore these beam normal SSA were predicted to be in the range of a few to hundred ppm for electron beam energies in the GeV range~\cite{Afanasev:2002gr, Gorchtein:2004ac, Pasquini:2004pv}. Although such asymmetries are small, the parity violation programs at the major electron laboratories have reached precisions on asymmetries with longitudinally polarized electron beams well  below the ppm level, and the next generation of such experiments is designed to reach precisions at the sub-ppb level~\cite{Kumar:2013yoa}. The beam normal SSA, 
which is due to TPE and thus parity conserving,  
has been measured over the past fifteen years as a spin-off by the parity-violation experimental collaborations at MIT-BATES (SAMPLE Coll.)~\cite{Wells:2000rx}, 
at MAMI (A4 Coll.)~\cite{Maas:2004pd, Rios:2017vsw}, and at JLab (G0 Coll.~\cite{Armstrong:2007vm, Androic:2011rh}, HAPPEX/PREX Coll.~\cite{Abrahamyan:2012cg}, and Qweak Coll.~\cite{Waidyawansa:2013yva}).  The measured beam normal SSA for the elastic 
$ep \to ep$ process ranges from a few ppm in the forward angular range to around a hundred ppm in the backward angular range, in good  agreement with theoretical TPE expectations.

Preliminary results from the QWeak Coll.~\cite{Nuruzzaman:2015vba,Dalton:2015lqa} 
for the beam normal SSA for the $e p \to e \Delta^+(1232)$ process indicate that the asymmetry for the inelastic process is  around an order of magnitude larger than the elastic asymmetry. It is the aim of this work to detail the formalism to understand this inelastic beam normal spin asymmetries and to study their sensitivity on the $\Delta(1232)$
electromagnetic FFs as well as on the $N^\ast \to \Delta$ electromagnetic transitions. 
 
The outline of this work is as follows. In Section II we briefly recall the definition of the beam normal SSA. 
In Section III, we describe the leading one-photon exchange amplitude to the 
$e p \to e \Delta$ process. Subsequently in Section IV, we give the general expression of the absorptive part of the two-photon exchange amplitude to the $e p \to e \Delta$ process, and describe the dominant regions in the phase space integrations. 
 In Section V, we provide the details of the model for the hadronic tensor entering the $e p \to e \Delta$ TPE amplitude which we use in this work. Besides the intermediate nucleon contribution, we subsequently describe the $\Delta(1232)$, $S_{11}(1535)$, and $D_{13}(1520)$ resonance intermediate state contributions. In Section VI, we show our results and discussions. We compare with the existing data for the Qweak@JLab experiment, and provide predictions for the A4@MAMI experiment. Our conclusions are given in Section VII. We provide the quark model relations to relate the electromagnetic $\Delta \to S_{11}$ and $\Delta \to D_{13}$ helicity amplitudes to the $N \to S_{11}$ and $N \to D_{13}$ helicity amplitudes in an Appendix.  

\section{Beam normal spin asymmetry }

The beam normal single spin asymmetry ($B_n$), corresponding with 
 the scattering of an electron with polarization {\it normal} to the scattering plane on a unpolarized proton target, 
is defined by~:
\begin{eqnarray}
B_n \,=\, 
\frac{\sigma_\uparrow-\sigma_\downarrow}{\sigma_\uparrow+\sigma_\downarrow}\,,
\label{eq:tasymm}
\end{eqnarray} 
where $\sigma_\uparrow$ ($\sigma_\downarrow$) denotes the cross section 
for an unpolarized target and for an electron spin 
parallel (anti-parallel) to the normal polarization vector, defined
as~:
\begin{eqnarray}
\xi^\mu \,=\, (\,0\,,\, \vec \xi \,), \quad \quad
\vec \xi \,\equiv \,  (\vec{k}\times\vec{k}') \,/\, | \vec{k}\times\vec{k}' | .
\label{eq:sn}
\end{eqnarray}
\indent
Applying the derivation of   
Ref.~\cite{DeRujula:1972te} to the case of a beam polarization normal to the scattering plane, 
$B_n$ can be expressed to order $e^2$ as:
\bea
B_n \simeq
\frac{ 2  \, {\rm Im} \left[\, \left( T_{1\gamma}\right)^\ast_{fi} \left( {\rm Abs} \,T_{2\gamma} \right)_{fi}  \right]}
{\sum \limits_{\rm spins} |T_{1\gamma}|^2} \, , 
\label{eq:an1}
\eea
where $T_{1\gamma}$ denotes the OPE amplitude, and ${\rm Abs}T_{2\gamma}$ the absorptive part of the TPE 
amplitude between the initial state $i$ and the final state $f$. The beam polarization in the initial state in Eq.~(\ref{eq:an1}) is 
understood along the direction of $\vec \xi$.  The numerator in Eq.~(\ref{eq:an1}) 
corresponds (to order $e^2$) to the difference of squared 
amplitudes for normal beam polarizations $\uparrow$ and 
$\downarrow$, while all other spins are summed over, whereas the denominator is the squared amplitude 
summed over all spins. 
The phase of the amplitude $T$ is defined through its relation to the S-matrix amplitude $S = 1- i \, T$.  
In Eq.~(\ref{eq:an1}), the absorptive part of the two-photon amplitude is defined 
as~\footnote{With this definition, one obtains the absorptive part from unitarity as: ${\rm Abs} \,T_{fi} = i \left[ (T)_{fi} - (T^\dagger)_{fi} \right]$. }:
\bea
\left( {\rm Abs} \,T_{2\gamma}\right)_{fi}  \equiv \sum \limits_n T_{nf}^\ast  \, T_{ni},
\eea
involving a sum over all physical (i.e. on-shell) intermediate states $n$. 

Generally, as illustrated by Eq.~\eqref{eq:an1}, one-photon exchange alone will give no beam normal single spin asymmetry.  The observed particle needs at least one further interaction.  When only the final electron is observed, which we consider in this work, this means two or more photon exchange.  In the resonance region, one can imagine observing instead a final pion, whence a non-zero $B_n$ is possible even for one-photon exchange~\cite{Buncher:2016nmv}, since the strong force guarantees final state interactions for the pion.
 
 In the following, we will evaluate Eq.~(\ref{eq:an1}) for the $e^- p \to e^- \Delta(1232)$ process. To this aim, 
 we will discuss in Section III the OPE amplitude $T_{1\gamma}$ , and in Sections IV and V the absorptive part of the 
 TPE amplitude. 

\section{One-photon exchange amplitude}

In this section, we briefly review the inelastic $e p \to e \Delta$ process in the one-photon exchange (OPE) approximation. 
The kinematics of the inelastic transition~: 
\begin{equation}
\label{eq:epedel}
e^-(k, s_e)+N(p, \lambda)\rightarrow e^-(k^\prime, s_e^\prime)+\Delta(p^\prime, \lambda^\prime),
\end{equation}
is described by four-vectors  $k (k^\prime)$ of the initial (final) electrons, 
and $p (p^\prime)$ of the nucleon ($\Delta$). 
Furthermore,  $s_e (s_e^\prime)$ denote the normal spin projections of the initial (final) electrons, 
and $\lambda (\lambda^\prime)$ the helicities of the nucleon ($\Delta$).
In this work, we will use the notation $q$ for the momentum transfer towards the hadronic system:
\begin{equation}
\label{eq:kin}
q=k-k^\prime=p^\prime-p,
\end{equation}
and adopt the usual definitions for the kinematical invariants of this process: 
\begin{equation}
\label{eq:qsqrnu}
s=(k+p)^2,\quad u =(k-p^\prime)^2, \quad t = q^2 \equiv - Q^2, 
\end{equation}
which are related as: $s + u - Q^2 = M_N^2 + M_\Delta^2 + 2 m_e^2$,
where $M_N (M_\Delta)$ are the nucleon ($\Delta$) masses respectively, 
and $m_e$ is the electron mass. 
Usually experiments are performed at fixed beam energy $E_e$, which determines $s$ as 
$s = M_N^2 + m_e^2 + 2 M_N E_e$. Furthermore, it is conventional in electron scattering to introduce the polarization parameter $\varepsilon$ of the virtual photon, which can be expressed in terms of the above kinematical invariants as (neglecting the electron mass):
\begin{equation}
\label{eq:eps}
\varepsilon =\frac{2(M_N^2M_\Delta^2-su)}{s^2+u^2-2M_N^2M_\Delta^2}.
\end{equation}

\begin{figure}[h]
\includegraphics[width=6cm]{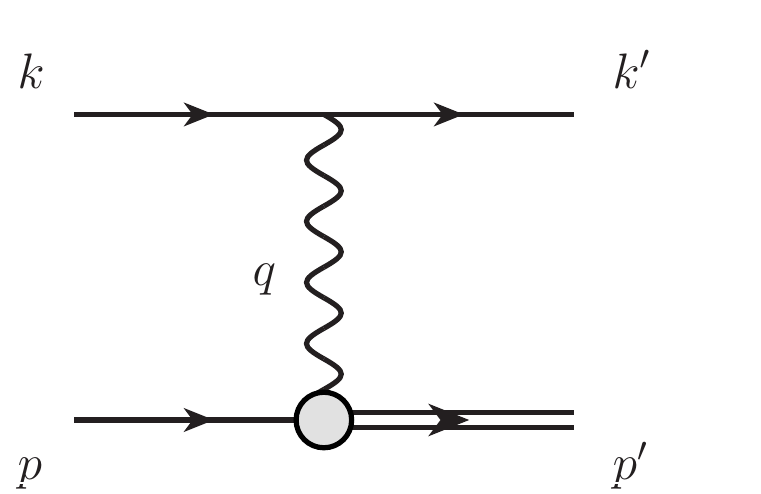}
\caption{The one-photon exchange diagram. 
The grey blob represents the electromagnetic vertex of the nucleon.}
\label{fig:gandeltreevertex}
\end{figure}

The OPE amplitude for the $e p \to e \Delta$ process is given by~\footnote{For simplicity of notation, we will redefine here and in the following of the paper the $T$-matrix elements by taken a global energy-momentum conservation factor $(2 \pi)^4 \delta^4(k + p - k^\prime - p^\prime)$ out of the $T$-matrix element.}:
\begin{eqnarray}
T_{1\gamma} &=& - \frac{e^2}{Q^2}\bar{u}(k^\prime,s_e^\prime) \gamma_\mu u(k,s_e) 
\langle \Delta(p^\prime, \lambda^\prime) | J^\mu(0) | N (p, \lambda)  \rangle, \nonumber \\
\label{eq:t1ga}
\end{eqnarray}
with $e$ the proton electric charge.
The matrix element of the hadronic current can be expressed in the covariant form~:
\begin{eqnarray}
\langle \Delta(p^\prime,  \lambda^\prime) | J^\mu(0) | N (p, \lambda) 
\rangle \equiv 
\bar u_\alpha (p^\prime, \lambda^\prime) 
\Gamma^{\alpha \mu}_{N \Delta}(p^\prime, p)
u(p,\lambda) , \nonumber \\ 
\eea
where $u$ is the nucleon spinor, and $u_\alpha$ is the Rarita-Schwinger spinor for the $\Delta$. 
Furthermore, the on-shell $\gamma^\ast N\Delta$ vertex is given by:
\begin{eqnarray}
\Gamma^{\alpha\mu}_{N \Delta}(p^\prime, p) &\equiv& \sqrt{\frac32}\frac{(M_\Delta+M_N)}{M_NQ_{N\Delta +}^2}
\left[ g_M(Q^2) i \, \varepsilon^{\alpha\mu\rho\sigma}p^\prime_\rho q_\sigma \nonumber \right.\\
&&\hspace{.75cm }\left.- g_E(Q^2)(q^\alpha p^{\prime \mu}-q\cdot p^\prime g^{\alpha\mu}) \gamma_5\right. 
\nonumber \\
&&\hspace{.75cm} \left.- g_C(Q^2)(q^\alpha q^\mu-q^2g^{\alpha\mu}) \gamma_5\right],
\label{eq:gaNDel}
\end{eqnarray}
where we use $\varepsilon_{0123} = +1$, and 
where $g_M$, $g_E$, and $g_C$ represent the three form factors (FFs) describing the $N \to \Delta$ vector 
transition~\cite{Pascalutsa:2006up}. 
We furthermore introduced the shorthand notation: 
\begin{equation}
Q_{N \Delta \pm}^2 \equiv Q^2+(M_\Delta\pm M_{N})^2. 
\label{eq:qpm}
\end{equation} 

Phenomenologically, the $\gamma^\ast N \Delta$ transition is usually expressed in terms of a different set of FFs introduced by  
Jones-Scadron~\cite{Jones:1972ky}, which are labeled 
$G_M^\ast$,  $G_E^\ast$, $G_C^\ast$, and describe the 
magnetic dipole (M1), electric quadrupole (E2), and Coulomb quadrupole (C2) transitions respectively. 
The latter have the property that they have a one-to-one relation with the 
imaginary parts of the pion electroproduction multipoles at the  resonance position, and have been 
extracted in experiment, see Ref.~\cite{Pascalutsa:2006up} for details. 
In terms of these Jones-Scadron FFs, the FFs entering Eq.~(\ref{eq:gaNDel}) are  
straightforwardly related as:   
\begin{eqnarray}
g_M&=&G^\ast_M-G^\ast_E,  \\
g_E&=&-\frac2{Q_{N\Delta -}^2}\left[(M_\Delta^2-M_N^2-Q^2)G_E^\ast+Q^2G_C^\ast\right], \nonumber \\
g_C&=&\frac1{Q_{N\Delta -}^2}\left[4M_\Delta^2 G_E^\ast-(M_\Delta^2-M_N^2-Q^2)G_C^\ast\right], \nonumber 
\end{eqnarray}
where all FFs are functions of $Q^2$.
The spin averaged squared matrix element for the $e p \to e \Delta$ process in the 
OPE approximation can then be expressed as~:  
\bea
 \sum_{\rm{spins}} |T_{1\gamma}|^2 \equiv \frac{e^4}{Q^2} \, D_{1 \gamma}(s, Q^2),
 \label{eq:T1gasqr}
\eea
where the function $D_{1\gamma}(s, Q^2)$ is given by: 
\bea
D_{1 \gamma}(s, Q^2) &=& \frac{2 \, Q_{N\Delta -}^2(M_\Delta+M_N)^2}{(1-\varepsilon) \, M_N^2} \nonumber \\
&\times&
\left[G_M^{\ast2}+3G_E^{\ast2}+\varepsilon\frac{Q^2}{M_\Delta^2}G_C^{\ast2}\right]. 
\label{eq:NDelOPE}
\eea
In this work, we will take the empirical information on the FFs $G_M^\ast(Q^2)$, 
$G_E^\ast(Q^2)$, and $G_C^\ast(Q^2)$, characterizing the electromagnetic  
$N \to \Delta$ transition, from the MAID2007 analysis~\cite{Drechsel:2007if,Tiator:2011pw}. 
In this analysis, the empirical $N \to \Delta$ transition FFs have been expressed as:
\bea
G_{M,E,C}^{\ast}(Q^2) = \left(\frac{Q_{N\Delta +}}{M_N + M_\Delta}\right) G_{M,E,C}^{\ast \, Ash}(Q^2), 
\eea
with the so-called Ash FFs $G_{M,E,C}^{\ast \, Ash}$ parameterized as~\cite{Drechsel:2007if,Tiator:2011pw}:
\bea
G_M^{\ast \, Ash}(Q^2) &=&  3.00 (1 + 0.01 Q^2) e^{-0.23 Q^2} G_D(Q^2), \nn \\
G_E^{\ast \, Ash}(Q^2) &=&  0.064 (1 - 0.021 Q^2) e^{-0.16 Q^2} G_D(Q^2), \nn \\
G_C^{\ast \, Ash}(Q^2) &=&  0.124 \frac{(1 + 0.120 Q^2)}{1 + 4.9 Q^2 / (4 M_N^2)} 
\left( \frac{4 M_\Delta^2}{M_\Delta^2 - M_N^2} \right) \nn \\
&\times& e^{-0.23 Q^2} G_D(Q^2), 
\eea
for $Q^2$ in GeV$^2$, and where $G_D(Q^2) = 1/(1 + Q^2 / 0.71)^2$ is the standard dipole FF.  
Note that the magnetic dipole $N \to \Delta$ transition provides by far the dominant contribution as 
$G_M^\ast(0) = 3.0$, whereas the electric and Coulomb quadrupole FFs are only at the few percent level relative to the magnetic dipole FF in the low $Q^2$ range. 

We like to notice that in the forward direction, $Q^2 \to 0$, the function $D_{1 \gamma}$ for the 
$ e p \to e \Delta$ process behaves, for fixed beam energy, approximately as:
\bea
&& D_{1 \gamma} \underset{Q^2 \to 0}{\longrightarrow}  \frac{4}{M_N^2} 
\left\{ \left( s - \frac{1}{2}(M_\Delta^2 + M_N^2) \right)^2 \right. \nn \\
&&\hspace{2.25cm}\left. + \frac{1}{4} (M_\Delta^2 - M_N^2)^2 \right\}  \left[ G_M^{\ast 2} + 3 G_E^{\ast 2} \right]. \quad
\label{eq:d1gainel}
\eea
In contrast, the corresponding function for the elastic process $e p \to e p$, which we denote by 
$D_{1 \gamma}^{el}$, behaves as~\cite{Pasquini:2004pv}:
\bea
&&D_{1 \gamma}^{el} \underset{Q^2 \to 0}{\longrightarrow}  
 \frac {16}{Q^2} (s - M_N^2)^2 F_1^2  \nonumber \\ 
&&\hspace{0.75cm} + \frac{4}{M_N^2} \left[ (s - M_N^2)^2  F_2^2 - 4 s M_N^2 F_1^2 \right] 
+ \mathcal{O}(Q^2)  , \quad 
\label{eq:d1gael}
\eea
where $F_1$ ($F_2$) are the Dirac (Pauli) FFs of the nucleon respectively. 
Eq.~(\ref{eq:d1gael}) then leads at forward angles to 
the characteristic $1/Q^4$ Rutherford behavior for the elastic  
OPE squared amplitude, defined by Eq.~(\ref{eq:T1gasqr}). 
On the other hand, the $e p \to e \Delta$ process, which necessarily involves a finite energy and momentum transfer, behaves as the Pauli ($F_2$) FF term of the elastic process, which only leads to a $1/Q^2$ behavior for the squared amplitude at small $Q^2$. We therefore see that the OPE cross section for the $e p \to e \Delta$ process, which enters the denominator of $B_n$, is suppressed by one power of $Q^2$ relative to its elastic counterpart. The TPE amplitude for the $e p \to e \Delta$ process, on the other hand, does not have this same suppression at forward angles, as we will see in the following.  
As $B_n$ is proportional to the TPE amplitude relative to the OPE amplitude, see Eq.~(\ref{eq:an1}), this leads to an enhancement of $B_n$ for the $e p \to e \Delta$ process at small values of $Q^2$, relative to its elastic counterpart.

\section{Imaginary (absorptive) part of the two-photon exchange amplitude}
\label{sec:impart}

In this section we relate the imaginary part of the
TPE amplitude, which appears in the numerator of $B_n$, to the absorptive part 
of the matrix element for the $e p\to e\Delta$ process, 
as shown in Fig.~\ref{fig:2gamma}. 
\begin{figure}[h]
\includegraphics[width=7cm]{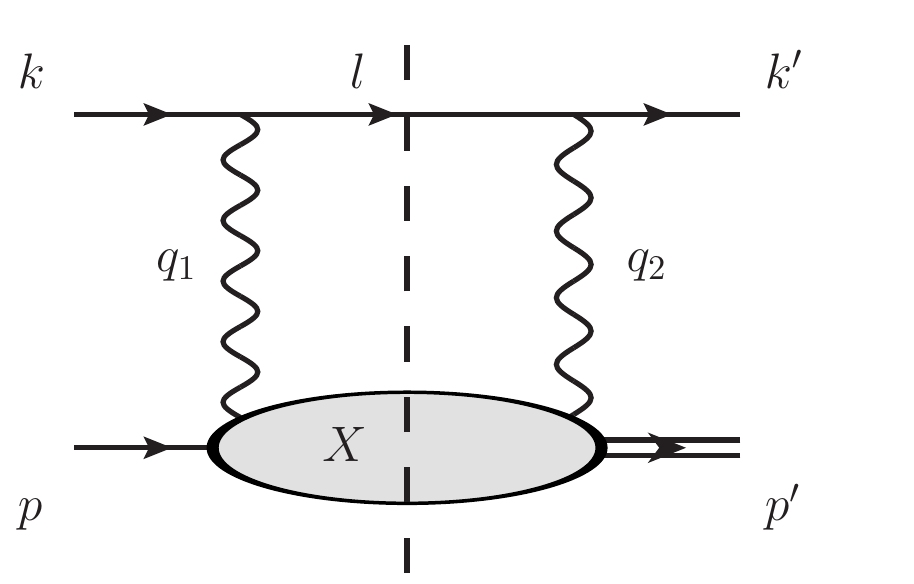}
\caption{The discontinuity of the two-photon exchange diagram. 
The cut blob represents the absorptive part of the doubly virtual Compton amplitude on a nucleon.}
\label{fig:2gamma}
\end{figure}

In the $e^- p$ {\it c.m.} frame, its contribution can be expressed as:
\bea
{\rm Abs} \, T_{2\gamma} &=& \int\frac{d^3\vec{l}}{(2\pi)^32E_{l}}
\bar{u}(k^\prime,s_e^\prime)\gamma_\mu(\gamma \cdot l+m_e)\gamma_\nu u(k,s_e) \nonumber \\
&&\hspace{1.5cm} \times \frac{e^4}{Q_1^2Q_2^2}\cdot W^{\mu\nu}(p^\prime,\lambda^\prime;p,\lambda) \,,
\label{eq:abs}
\eea
\noindent
where the momenta are defined as indicated on Fig.~\ref{fig:2gamma}, 
with $q_1 \equiv k - l$, $q_2 \equiv k^\prime - l$, $q_1 - q_2 = q$, and where 
$E_l$ is the energy of the intermediate lepton.
Furthermore, $Q_1^2 \equiv -q_1^2 = - (k - l)^2$ and 
$Q_2^2 \equiv -q_2^2 = - (k^\prime - l)^2$ correspond with the 
virtualities of the two spacelike photons. 
Denoting the {\it c.m.} angle between 
initial and final electrons as $\theta_{cm}$,
the momentum transfer $Q^2 \equiv - q^2 > 0$ can be expressed as~:
\begin{eqnarray}
Q^2 = \frac{(s - M_N^2)(s - M_\Delta^2)}{2 \, s} \, \left( 1 - \cos \theta_{cm} \right) 
\,+\, {\mathcal{O}}(m_e^2). 
\end{eqnarray}

In Eq.~(\ref{eq:abs}), the 
hadronic tensor $W^{\mu\nu}(p',\lambda^\prime;p,\lambda)$ 
corresponds with the absorptive part
of the doubly virtual $\gamma^\ast N\to \gamma^\ast \Delta$ tensor 
for two {\it space-like} photons~:
\bea
&&W^{\mu\nu}(p^\prime,\lambda^\prime;p,\lambda) =
\sum_X \,(2\pi)^4 \, \delta^4(p+q_1-p_X)  \nonumber \\
&&\hspace{1.cm}\times
\langle \Delta(p^\prime, \lambda^\prime) |J^{\dagger \mu}(0)|X\left> \, \right<X| J^\nu(0)| N(p,  \lambda) \rangle , \quad \quad 
\label{eq:wtensor}
\eea
where the sum goes over all possible {\it on-shell} intermediate hadronic 
states $X$. 
We will use the unitarity relation to express the full non-forward tensor 
in terms of electroproduction amplitudes $\gamma^* N \to X$. 
The number of intermediate states $X$ which one considers in the 
calculation will then put a limit on how high in energy one can 
reliably calculate the hadronic tensor of Eq.~(\ref{eq:wtensor}). 
In this work, we will model the tensor $W^{\mu\nu}$ as 
a sum over different baryon intermediate states, and  
will explicitly consider $X = N$,  $\Delta(1232)$, $S_{11}(1535)$, and 
$D_{13}(1520)$, resonance contributions. 

The phase space integral in Eq.~(\ref{eq:abs}) runs over 
the 3-momentum of the intermediate (on-shell) electron. 
Evaluating the process in the $e^- p$ {\it c.m.} system, 
we can express the {\it c.m.} momentum of the intermediate electron as~:
\begin{equation}
|\vec{l}|^2 = \frac{1}{4s} \left[ (\sqrt{s} - m_e)^2 - W^2 \right]   \left[ (\sqrt{s} + m_e)^2 - W^2 \right] ,
\label{eq:kink1} 
\end{equation}
where $W^2 \equiv p_X^2$ is 
the squared invariant mass of the intermediate state
$X$. 
The {\it c.m.} momenta of the initial (and final) electrons are  
given by the analogous expression as Eq.~(\ref{eq:kink1}) by replacing 
$W^2$ with $M_N^2$ $(M_\Delta^2)$ respectively. 
The phase space integral in Eq.~(\ref{eq:abs}) 
depends, besides the magnitude $|\vec l|$,   
upon the solid angle of the intermediate electron. 
We define the polar {\it c.m.} angle $\theta_1$ of the intermediate electron 
w.r.t. to the direction of the initial electron. The azimuthal angle 
$\phi_1$ is chosen such that $\phi_1 = 0$ 
corresponds with the scattering plane of the $e p \to e \Delta$ process. 
Having defined the kinematics of the intermediate electron, we can 
express the virtuality of both exchanged photons. The virtuality of 
the photon with four-momentum $q_1$ is given by~:
\begin{multline}
Q_1^2=\frac1{2s}\left[(s-M_N^2 + m_e^2)(s-W^2 + m_e^2) - 4 m_e^2 s \right.\\
\left.-\sqrt{(s-M_N^2+m_e^2)^2-4m_e^2s }\right.\\
\left.\times\sqrt{(s-W^2+m_e^2)^2-4m_e^2s}\cos\theta_1\right].
\label{eq:q1virt} 
\end{multline}
The virtuality $Q_2^2$ of the second photon has an analogous expression 
as Eq.~(\ref{eq:q1virt}) with the replacements $M_N \to M_\Delta$ and $\cos \theta_1 \to \cos \theta_2$, 
where $\theta_2$ is the angle between the intermediate and final electrons. 
In terms of the polar and azimuthal angles $\theta_1$ and $\phi_1$ of the 
intermediate electron, one can express~:
\begin{eqnarray}
\cos \theta_2 \,=\, \sin \theta_{cm} \, \sin \theta_1 \, \cos \phi_1 
\,+\, \cos \theta_{cm} \, \cos \theta_1 .
\end{eqnarray}
\indent
In case the intermediate electron 
is collinear with the initial electron (i.e. for 
$\theta_1 \to 0$, $\phi_1 \to 0$), denoting the virtual photon virtualities for this kinematical situation by 
$Q_{i, VCS}^2 \equiv  Q_i^2 (\theta_1 = 0, \phi_1 = 0)$,  
one obtains from Eq.~(\ref{eq:q1virt}) that:
\begin{eqnarray}
Q_{1, \, VCS}^2 &\simeq& m_e^2  \frac{(W^2 - M_N^2)^2}{(s - W^2)(s - M_N^2)}, \nonumber \\ 
Q_{2, \, VCS}^2 &\simeq& \frac{(s - W^2)}{(s - M_N^2)} Q^2 + {\mathcal{O}}(m_e^2). 
\label{eq:quasivcs}
\end{eqnarray}
We thus see that when the intermediate and initial electrons are collinear, 
the photon with momentum $\vec q_1 = \vec k - \vec k_1$ is 
also collinear with this direction, and its virtuality becomes of order of ${\mathcal{O}}(m_e^2)$, whereas 
the other photon has a large virtuality, of order $Q^2$. 
For the case $W = M_N$, this precisely corresponds with 
the situation where the first photon is soft (i.e. $q_1 \to 0$), and 
where the second photon carries the full momentum transfer 
$Q_2^2 \simeq Q^2$. 
For the case $W > M_N$, the first photon is 
hard but becomes quasi-real (i.e. $Q_1^2 \sim m_e^2$). 
In this case, the virtuality of the second photon is smaller than $Q^2$. 
An analogous situation occurs when the intermediate electron is 
collinear with the final electron 
(i.e. $\theta_2 \to 0$, $\phi_1 \to 0$, which is equivalent with 
$\theta_1 \to \theta_{cm}$). The corresponding photon virtualities are obtained from 
Eq.~(\ref{eq:quasivcs}) by the replacements $Q_{1, \, VCS}^2 \leftrightarrow Q_{2, \, VCS}^2$ and 
$M_N \leftrightarrow M_\Delta$. The second photon is quasi-real in this case, and the first photon carries a virtuality smaller than $Q^2$. For the special case of a $\Delta$ intermediate state $W = M_\Delta$, the second photon becomes soft, and the first photon carries the full momentum transfer $Q^2$.   
These phase space regions with one quasi-real photon and one virtual photon 
correspond with quasi virtual Compton scattering (quasi-VCS), 
and correspond at the lepton side with the Bethe-Heitler process, see  
e.g. Ref.~\cite{Guichon:1998xv}  for details.
They lead to large enhancements in the integrand entering the absorptive part of the TPE amplitude.   

Besides the near singularities corresponding with quasi-VCS, where  
the intermediate electron is collinear with either the incoming or outgoing electrons, 
the TPE process also has a near singularity when 
the intermediate electron momentum goes to zero $|\vec l | \to 0$ 
(i.e. the intermediate electron is soft). 
In this case the first photon takes on the full momentum of the 
initial electron, i.e. $\vec q_1 \to \vec k$, whereas the 
second photon takes on the full momentum of the final electron, 
i.e. $\vec q_2 \to \vec k^\prime$. 
One immediately sees from Eq.~(\ref{eq:kink1}) 
that this situation occurs when the invariant mass of the hadronic 
state takes on its maximal value $W = W_{max} \equiv \sqrt{s} - m_e$.   
In this case, the photon virtualities are given by~:
\begin{eqnarray}
Q_{1, \, RCS}^2  &=& \frac{m_e}{\sqrt{s}} \left\{ \left(\sqrt{s} - m_e \right)^2 - M_N^2 \right\}, \nn \\
Q_{2, \, RCS}^2  &=& \frac{m_e}{\sqrt{s}} \left\{ \left(\sqrt{s} - m_e \right)^2 - M_\Delta^2 \right\} . 
\label{eq:q1real}
\end{eqnarray}
This kinematical situation with two quasi-real photons, corresponding with 
quasi-real Compton scattering (quasi-RCS), also leads to an enhancement in the corresponding 
integrand of ${\rm Abs} T_{2\gamma} $.   

\begin{figure}
\includegraphics[width=7.cm]{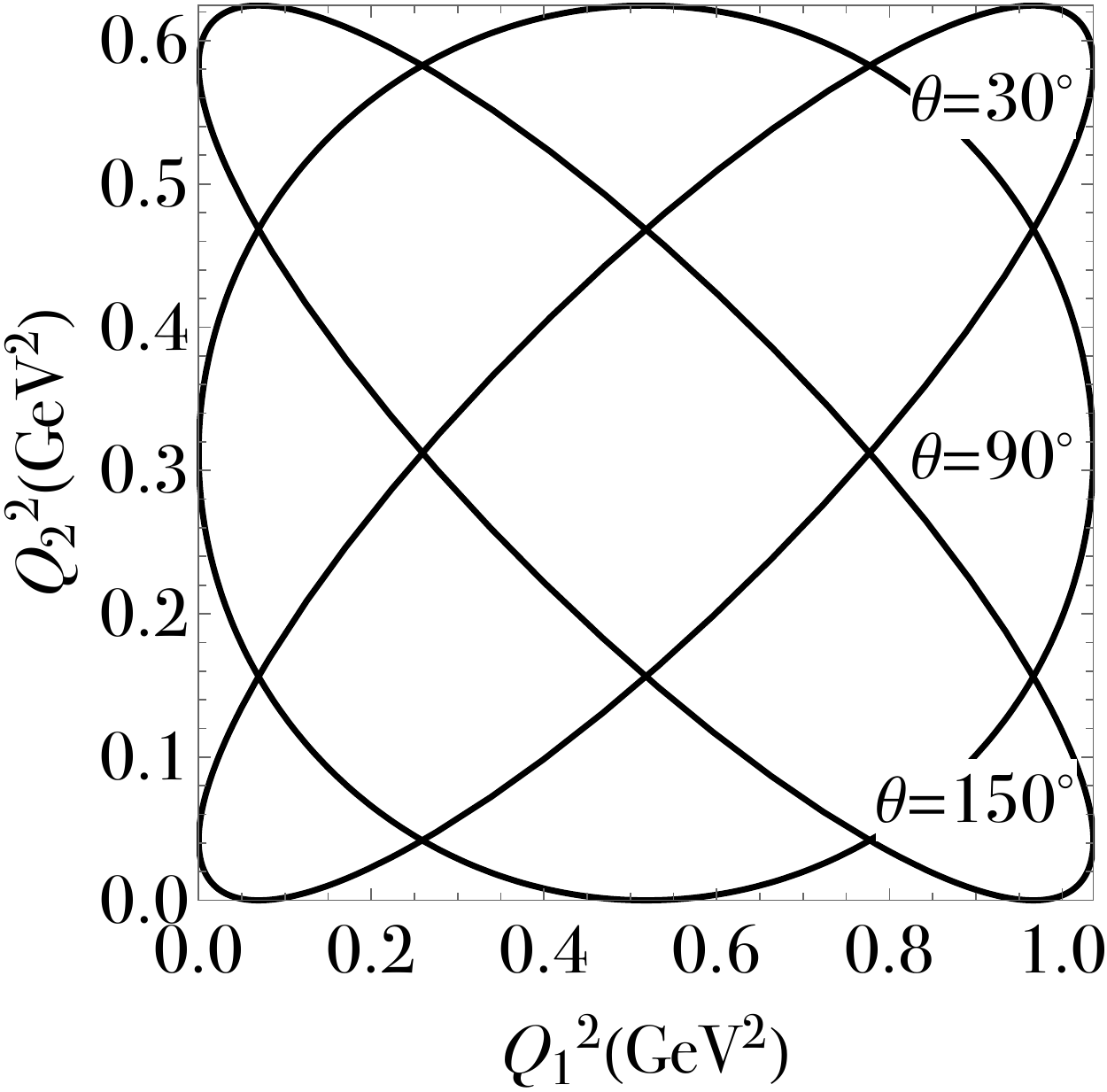}\\
\includegraphics[width=7.cm]{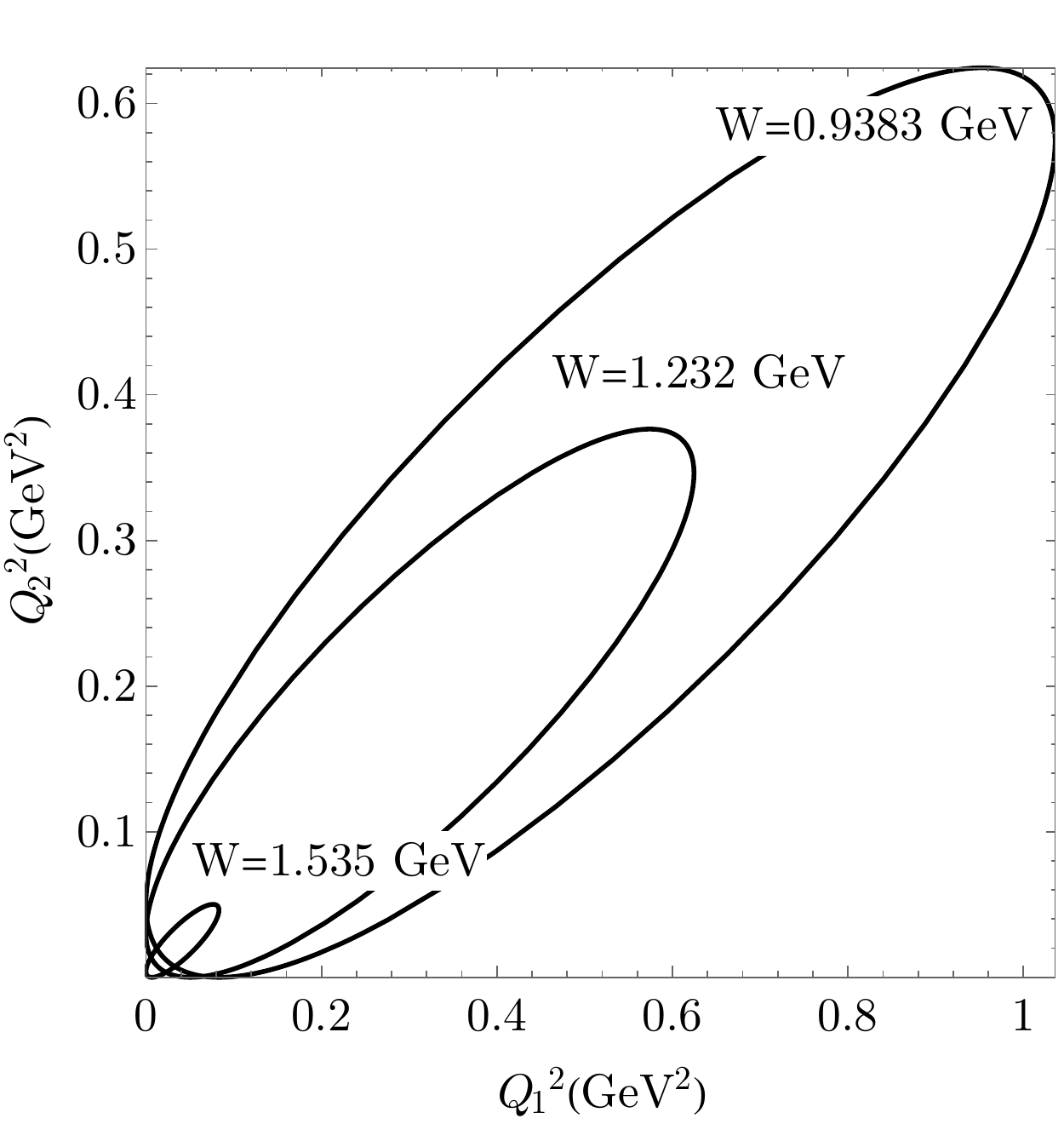}
\caption{Kinematical accessible region for the virtualities
$Q_1^2, Q_2^2$ in the phase space integral of 
Eqs.~(\ref{eq:abs}, \ref{eq:an2}) entering the $e p \to e \Delta$ process.  
The upper panel shows the phase space regions for different 
{\it c.m.} angles $\theta_{cm}$ as indicated on the ellipses for $E_e = 0.855$ GeV ($s=2.485\;\mathrm{GeV}^2$),  
and for $W=0.9383\;\mathrm{GeV}$ (i.e. for a nucleon intermediate state). The lower panel shows the allowed values of the 
photon virtualities for different intermediate states for $\theta_{cm} = 30^o$. 
We show three cases corresponding with the contribution of $N$, $\Delta(1232)$ and $S_{11}(1535)$ excitations.
The accessible regions correspond with the interior of the ellipses. 
The intersection with the axes correspond with quasi-VCS, whereas  
the situation at $W = \sqrt{s} - m_e$ where all ellipses shrink to the point 
$Q_1^2 = Q_2^2 \simeq 0$ corresponds with quasi-RCS.}
\label{fig:qsqr_bounds}
\end{figure} 

In the upper panel of Fig.~\ref{fig:qsqr_bounds}, we show the 
kinematical accessible regions for the virtualities
$Q_1^2, Q_2^2$ in the phase space integral of Eq.~(\ref{eq:abs}) for a beam energy of $E_e = 0.855$~GeV corresponding with the A4@MAMI experiment, for different values of the {\it c.m.} angle $\theta_{cm}$.  In the lower panel we display these phase space regions for three different values of $W$, corresponding with the $N$, $\Delta(1232)$, and $S_{11}(1535)$ intermediate states.  
We notice from Fig.~\ref{fig:qsqr_bounds} that the largest possible photon virtualities in the TPE amplitude occur for the nucleon intermediate state, whereas for the $S_{11}(1535)$ intermediate state both photons have very small virtualities.

Using Eq.~(\ref{eq:abs}) for the absorptive part of the TPE amplitude, we can then express the normal spin asymmetry $B_n$ of Eq.~(\ref{eq:an1})  
for the $e p \to e \Delta$ process in terms of a
3-dimensional phase-space integral:
\bea
B_n &=& - \frac{e^2}{D_{1 \gamma}(s,Q^2)} \, \frac{1}{(2\pi)^3}
\int_{M^2}^{(\sqrt{s}-m_e)^2}dW^2\, \left( \frac{s - W^2}{8 \, s } \right) \nonumber \\
&\times& \int d\Omega_{1}\frac{1}{Q_1^2 \, Q_2^2}\,{\rm Im}
\left( L_{\kappa\mu\nu} \, H^{\kappa\mu\nu}\right)\,,
\label{eq:an2}
\eea
where the denominator factor $D_{1 \gamma}(s, Q^2)$  is originating from the OPE process as given by Eq.~(\ref{eq:NDelOPE}), and $d \Omega_1 = d \cos \theta_1 d \phi_1$.  

Equivalently, the phase space integration in Eq.~(\ref{eq:an2}) can be re-expressed in a Lorentz invariant way as an integral over photon virtualities $Q_1^2$ and $Q_2^2$ by using the Jacobian
\bea
J=\left|\frac{\partial Q_1^2}{\partial \cos\theta_1}\frac{\partial Q_2^2}{\partial \phi_1}\right| .
\eea
Using Eq.(\ref{eq:q1virt}) and an analogous expression for $Q_2^2$, the Jacobian is given by 
\begin{eqnarray}
J &=& \left[(s-W^2+m_e^2)^2-4m_e^2s\right] / (4s^2) \nn \\
&\times& \left[(s-M_N^2+m_e^2)^2-4m_e^2s \right]^{1/2} \nn \\
&\times& \left[ (s-M_\Delta^2+m_e^2)^2-4m_e^2s \right]^{1/2}  \nn \\
&\times& \sin\theta_{cm}\sin\theta_1\sin\phi_1,
\end{eqnarray}
leading to the equivalent expression for $B_n$:
\bea
B_n &=& - \frac{e^2}{D_{1 \gamma}(s,Q^2)} \, \frac{1}{(2\pi)^3}
\int_{M^2}^{(\sqrt{s}-m_e)^2}dW^2\, \left( \frac{s - W^2}{8 \, s } \right) \nonumber \\
&\times& \int \mathrm{d}Q_{1}^2\mathrm{d}Q_{2}^2\frac{J^{-1}(Q_1,Q_2)}{Q_1^2 \, Q_2^2}{\rm Im}
\left( L_{\kappa\mu\nu} \, H^{\kappa\mu\nu}\right)\, ,
\label{eq:an3}
\eea
where the $(Q_1^2, Q_2^2)$ integration regions cover the inside of ellipses as displayed e.g. in Fig.~\ref{fig:qsqr_bounds}. 

The integrand in Eqs.~(\ref{eq:an2}, \ref{eq:an3}) arising from the 
interference between the OPE and TPE amplitudes has been expressed as a product of a 
lepton tensor $L_{\lambda \mu \nu}$ and a hadron tensor $H^{\lambda \mu \nu}$. 
The polarized lepton tensor can be expressed as a trace using the spin projection technique:
\begin{equation}
 L_{\kappa\mu\nu}\hspace{-0.03cm} = \hspace{-0.03cm} \mathrm{Tr} \hspace{-0.03cm} 
\left\{ \gamma_\kappa  (\sl{k'} + m_e) \gamma_\mu (\sl{\, l} + m_e) \gamma_\nu 
\gamma_5 \, \sl{\xi} ( \sl{k} + m_e) \right\}\hspace{-0.03cm} , \hspace{-0.07cm} 
\label{eq:lept2}
\end{equation}
where $\xi^\alpha$ is the polarization vector of Eq.~(\ref{eq:sn}) for an electron polarized 
normal to the scattering plane.  
We see from Eq.~(\ref{eq:lept2}) that the polarized lepton tensor 
vanishes for massless electrons. Keeping only the leading term in $m_e$, it is 
given by:
\beqn
L_{\kappa\mu\nu}&=& m_e \left( \,  
- \mathrm{Tr} 
\left\{ \gamma_5 \gamma_\mu \sl{\, l} \gamma_\nu \, \sl{\xi} \, \sl{k} \,  
\gamma_\kappa \right\}
+ \mathrm{Tr} 
\left\{ \gamma_5 \sl{k'}  \gamma_\mu \sl{\, l}  \gamma_\nu \, \sl{\xi}  \,
\gamma_\kappa \right\} \right. \nonumber \\
&& \hspace{.75cm} - \left.  \mathrm{Tr} 
\left\{ \gamma_5 \sl{k'}  \gamma_\mu  \gamma_\nu \, \sl{\xi} \, \sl{k} \,  
\gamma_\kappa \right\} 
\, \right)
+  \, {\mathcal{O}}(m_e^2) .
\label{eq:lept3}
\eeqn
Furthermore, the unpolarized hadron tensor $H^{\lambda \mu \nu}$ is given by
\bea
H^{\kappa\mu\nu}&=& \sum \limits_{\lambda, \lambda^\prime} 
\left[\bar{u}_\alpha(p',\lambda^\prime)\Gamma^{\alpha \kappa}_{N \Delta}(p', p) u(p,\lambda)\right]^* \nonumber \\
&&\hspace{0.75cm}\times W^{\mu\nu}(p',\lambda^\prime;p,\lambda)  .
\label{eq:hadt}
\eea
We can express the sum over the hadron spins in Eq.~(\ref{eq:hadt}) 
as a trace by expressing the hadron tensor $W^{\mu \nu}$ through an operator $\hat{W}$ in spin space, defined as:
\begin{equation}
W^{\mu\nu}(p',\lambda^\prime;p,\lambda) \equiv 
\bar{u}_\beta(p',\lambda^\prime)\hat{W}^{\beta\mu\nu}(p',p)u(p,\lambda). 
\label{eq:What}
\end{equation} 
The spin summation in Eq.~(\ref{eq:hadt}) can then be worked out as:
\bea
H^{\kappa\mu\nu} &=&  \mathrm{Tr} \left\{  \tilde \Gamma^{\alpha\kappa}_{N \Delta} (p', p)  P^{(3/2)}_{\alpha \beta}(p',M_\Delta)\hat{W}^{\beta \mu\nu}(p^\prime,p) \right. \nn \\
&&\hspace{0.5cm} \times \left. P^{(1/2)}(p,M_N) \right\} , 
\eea
where $\tilde \Gamma^{\alpha \beta}_{N \Delta} \equiv \gamma^0 
\left(\Gamma^{\alpha \beta}_{N \Delta}\right)^\dagger \gamma^0$ stands for the adjoint operator, and 
where the spin-3/2 and spin-1/2 projectors for a state of mass $M$ are defined by:
\begin{eqnarray}
P^{(1/2)}(p,M)&=&\sl{p}+M,  \\
P^{(3/2)}_{\alpha \beta}(p,M)&=&(\sl{p}+M)
\left[-g_{\alpha \beta} +\frac{1}{3}\gamma_\alpha \gamma_\beta \right. \nonumber\\ 
&&\left. \hspace{1cm}
+\frac1{3p^2}(\sl{p}\gamma_\alpha p_\beta+p_\alpha \gamma_\beta \sl{p})\right].
\end{eqnarray}

For narrow intermediate states $X$, which we will consider in the following, the hadronic tensor is given by~:
\bea
H^{\kappa \mu \nu} \equiv (2 \pi) \delta(W^2 - M_X^2) \tilde H^{\kappa \mu \nu},
\eea
which then reduces the expression for $B_n$ in Eq.~(\ref{eq:an3}) to a 2-dimensional integral:
\bea
B_n &=& - \frac{1}{(2\pi)^2}\frac{e^2}{D_{1 \gamma}(s,Q^2)} \, 
\left( \frac{s - M_X^2}{8 \, s } \right)  \theta(s - M_X^2) \nonumber \\
&\times& \int \mathrm{d}Q_{1}^2\mathrm{d}Q_{2}^2\frac{J^{-1}(Q_1,Q_2)}{Q_1^2 \, Q_2^2}{\rm Im}
\left( L_{\kappa\mu\nu} \, \tilde H^{\kappa\mu\nu}\right)\,.
\label{eq:an4}
\eea

\section{Models for the hadronic tensor}
\label{sec:models}

In this Section, we will model the hadronic tensor $\hat W^{\beta \mu\nu}$ of Eq.~(\ref{eq:What}) as 
a sum over different baryon intermediate states. 
We will explicitly consider $X = N$,  $\Delta(1232)$, $S_{11}(1535)$, and 
$D_{13}(1520)$ resonance contributions in the blob of Fig.~\ref{fig:2gamma}. 
The  nucleon contribution is calculable based on the empirical electromagnetic FFs for the nucleon and 
for the $N \to \Delta$ transition. 
We will express the $\Delta$ intermediate state contribution in terms of the $\Delta$ electromagnetic FFs, and will use a lattice calculation for the latter for an estimate. 
To estimate the unknown $\Delta \to S_{11} $ and $\Delta \to D_{13}$ electromagnetic transitions, we will use a constituent quark model to relate them to the corresponding FFs for the $N \to S_{11}$ and $N \to D_{13}$ electromagnetic transitions. The latter FFs  will be taken from experiment. 
We will detail these different contributions in the following.  

\subsection{Nucleon intermediate state contribution}

The contribution to $\hat W^{\beta \mu \nu}$,
corresponding with the nucleon  
intermediate state in Fig.~\ref{fig:2gamma},
is exactly calculable in terms of on-shell $\gamma^\ast NN$  and  $\gamma^\ast N\Delta$ vertices as:
\beqn
\hat{W}^{\beta \mu \nu}_{N}(p^\prime,p) \hspace{-0.1cm}
&=& 2\pi \, \delta(W^2-M_N^2) \; \Gamma^{\beta \mu}_{N \Delta}(p^\prime, p_N)  \nonumber \\
&\times&  P^{(1/2)}(p_N,M_N)  
 \Gamma^\nu_{NN}(p_N, p)  , \quad
\label{eq:welast}
\eeqn
with $p_N \equiv p + q_1$, where $\Gamma^{\beta \mu}_{N \Delta}$ is as in Eq.~(\ref{eq:gaNDel}), 
and the on-shell $\gamma^\ast NN$ vertex 
$\Gamma^\nu_{N N}$ is given by:
\begin{equation}
\Gamma^\nu_{N N} (p_N,p)\equiv (F_1+F_2)\gamma^\nu-F_2\frac{(p+p_N)^\nu}{2M_N},
\end{equation}
with $F_1 (F_2)$ the Dirac (Pauli) proton FFs respectively. 
For the nucleon intermediate state contribution, the unpolarized hadronic tensor 
entering Eqs.~(\ref{eq:an2}, \ref{eq:an3}) for $B_n$ can be written as:
\begin{align}
H^{\kappa\mu\nu}_N &=  2\pi \, \delta(W^2-M_N^2) \, \nonumber \\
&\times \mathrm{Tr} \left\{\tilde \Gamma^{\alpha\kappa}_{N \Delta}(p^\prime, p) \, P^{(3/2)}_{\alpha \beta}(p^\prime,M_\Delta) \,
\Gamma^{\beta \mu}_{N \Delta}(p', p_N) \right. \nonumber \\ 
&\left. \times P^{(1/2)}(p_N,M_N) \Gamma^\nu_{N N}(p_N, p) 
P^{(1/2)}(p,M_N) \right\} . 
\end{align}

\subsection{$\Delta(1232)$ intermediate state contribution}

The matrix element of the electromagnetic current operator $J^\mu$ 
between spin 3/2 states can be decomposed into four multipole transitions:
a Coulomb monopole (E0), a magnetic dipole (M1), 
an electric quadrupole (E2) and 
a magnetic octupole (M3).
We firstly write a Lorentz-covariant decomposition for the 
on-shell $\gamma^* \Delta \Delta$ vertex which exhibits
manifest electromagnetic gauge-invariance as~\cite{Pascalutsa:2006up}: 
\bea
\langle \Delta(p^\prime,  \lambda^\prime) | J^\mu(0) | \Delta (p, \lambda) 
\rangle &\equiv &
\bar u_\alpha (p^\prime, \lambda^\prime) 
\Gamma^{\alpha \beta \mu}_{\Delta \Delta}(p^\prime, p)
u_\beta(p,\lambda) , \nn \\
 &&
\eea
where $\lambda$ ($\lambda^\prime$) are the initial (final) $\Delta$ helicities, and where
$\Gamma^{\alpha \beta \mu}_{\Delta \Delta}$ is given by:  
\bea 
\Gamma^{\alpha \beta \mu}_{\Delta \Delta}(p^\prime, p) &=& - \left[  
F^\Delta_1  g^{\alpha \beta}  
+ F^\Delta_3 \frac{q^\alpha q^\beta}{(2 M_\Delta)^2} 
\right] \gamma^\mu  \nonumber \\
&-& \left[ F^\Delta_2  g^{\alpha \beta}
+ F^\Delta_4 \frac{q^\alpha q^\beta}{(2M_\Delta)^2}\right] 
\frac{i \sigma^{\mu\nu} q_\nu}{2 M_\Delta} , \quad 
\label{eq:gadeldeltree}
\end{eqnarray}
where $q = p^\prime - p$.
$F^\Delta_{1,2,3,4}$ are the $\Delta$ electromagnetic FFs and depend on $Q^2$. 
Note that $F^\Delta_1(0) = e_\Delta$ is the $\Delta$ electric charge in units of $e$ 
(e.g., $e_{\Delta^+} = +1$). 
For further use we also define the quantity 
$\tau_\Delta \equiv Q^2 / (4 M_\Delta^2)$. 

A physical interpretation of the four electromagnetic $\Delta \to \Delta$ 
transitions can be obtained by performing a multipole 
decomposition~\cite{Weber:1978dh,Nozawa:1990gt}. 
The FFs $F^\Delta_{1,2,3,4}$ can be expressed 
in terms of the multipole form factors $G_{E0}$, $G_{M1}$, $G_{E2}$, and $G_{M3}$, as~\cite{Alexandrou:2009hs}: 
\begin{eqnarray}
F^\Delta_1 &=& \frac{1}{1 + \tau_\Delta} 
\left\{ G^\Delta_{E0} - \frac{2 \tau_\Delta}{3} G^\Delta_{E2} \right.  \nonumber \\
&&\left. \hspace{1.5cm} + \tau_\Delta \Big[ G^\Delta_{M1} - \frac{4 \tau_\Delta}{5} G^\Delta_{M3} \Big] \right\}, 
\nonumber \\
F^\Delta_2 &=& -\frac{1}{1 + \tau_\Delta} 
\left\{ G^\Delta_{E0} - \frac{2 \tau_\Delta}{3} G^\Delta_{E2}   
- G^\Delta_{M1} + \frac{4\tau_\Delta}{5} G^\Delta_{M3}  \right\},  \nn \\
F^\Delta_3 &=& \frac{2}{(1 + \tau_\Delta)^2}
\left\{ G^\Delta_{E0} - \left( 1 + \frac{2 \tau_\Delta}{3} \right)  G^\Delta_{E2}  \right. \nonumber \\
&&\left. \hspace{1.5cm}+ \tau_\Delta \Big[ G^\Delta_{M1} - \left( 1 + \frac{4 \tau_\Delta}{5} \right) G^\Delta_{M3} \Big] 
\right\}, \nonumber \\
F^\Delta_4 &=& - \frac{2}{(1 + \tau_\Delta)^2} 
\left\{ G^\Delta_{E0} - \left( 1 + \frac{2 \tau_\Delta}{3} \right)  G^\Delta_{E2}  \right. \nonumber \\
&&\left. \hspace{1.5cm} -  
\Big[ G^\Delta_{M1} - \left( 1 + \frac{4 \tau_\Delta}{5} \right) G^\Delta_{M3} \Big] 
\right\}. 
\label{eq:deldelFF}
\end{eqnarray}
At $Q^2 = 0$, the multipole FFs define 
the charge $e_\Delta$, the magnetic dipole moment $\mu_\Delta$,
the electric quadrupole moment $\mathcal{Q}_\Delta$,  
and the magnetic octupole moment $\mathcal{O}_\Delta$ as:
\begin{eqnarray}
\label{eq:EMmoments}
e_\Delta = G^\Delta_{E0}(0) , \quad \quad 
&&\mu_\Delta = \frac{e}{2 M_\Delta} G^\Delta_{M1}(0) , \nonumber \\
\mathcal{Q}_\Delta =  \frac{ e}{ M_\Delta^2} G^\Delta_{E2}(0) , \quad \quad  
&& \mathcal{O}_\Delta =   \frac{e}{2M_\Delta^3} G^\Delta_{M3}(0) .
\end{eqnarray}

The inelastic contribution to $\hat W^{\beta \mu \nu}$,
corresponding with the $\Delta(1232)$   
intermediate state in the blob of Fig.~\ref{fig:2gamma},
is exactly calculable in terms of on-shell $\gamma^\ast N \Delta$ and $\gamma^\ast \Delta \Delta$ 
electromagnetic vertices as:
\bea
\hat{W}^{\beta \mu \nu}_{\Delta}(p^\prime,p) 
&=& 2\pi \, \delta(W^2-M_\Delta^2) \; \Gamma^{\beta \gamma \mu}_{\Delta \Delta}(p^\prime, p_\Delta)   \nonumber \\
&\times&    P^{(3/2)}_{\gamma \delta }(p_\Delta, M_\Delta) 
 \Gamma^{\delta \nu}_{N \Delta}(p_\Delta, p)  . \quad 
\label{eq:wdelta}
\eea
with $p_\Delta \equiv p + q_1$.
This allows us, for the $\Delta$ intermediate state contribution, to evaluate the unpolarized hadronic tensor 
entering Eqs.~(\ref{eq:an2}, \ref{eq:an3}) for $B_n$ as:
\bea
H^{\kappa\mu\nu}_\Delta &=&  2\pi \, \delta(W^2-M_\Delta^2) \, \nonumber \\
&\times& \mathrm{Tr} \left\{\tilde \Gamma^{\alpha\kappa}_{N \Delta}(p^\prime, p)  P^{(3/2)}_{\alpha \beta}(p^\prime,M_\Delta) 
\Gamma^{\beta \gamma \mu}_{\Delta \Delta}(p^\prime, p_\Delta) \right. \nonumber \\ 
&\times&  \left. P^{(3/2)}_{\gamma \delta }(p_\Delta,M_\Delta) 
 \Gamma^{\delta \nu}_{N \Delta}(p_\Delta, p) P^{(1/2)}(p,M_N) \right\}. \quad \;\;
\eea

In the following, we will study the sensitivity of $B_n$ to the $\Delta$ electromagnetic FFs. 
For the purpose of obtaining an estimate on the expected size of $B_n$, 
we will also directly compare with lattice calculations for the $\Delta$ FFs. 
We will use the results for the hybrid lattice calculation of Ref.~\cite{Alexandrou:2009hs}, which was performed for a pion mass of 
$m_\pi = 353$~MeV.  
The lattice results for $G^\Delta_{E0}$, were fitted in Ref.~\cite{Alexandrou:2009hs} by a  dipole parameterization:
\begin{eqnarray} 
G^\Delta_{E0}(Q^2) &=& \frac{1}{\left( 1 + Q^2 / \Lambda_{E0}^2 \right)^2} , 
\label{eq:ge0lattice}
\end{eqnarray}
with resulting fit value~:
\bea 
\Lambda_{E0}^2 = 1.160 \pm 0.078~\mathrm{GeV}^2.   
\eea
The FFs $G^\Delta_{M1}$ and $G^\Delta_{E2}$, were fitted by exponential parameterizations 
 since the expected large $Q^2$-dependence for these FFs
drops stronger than a dipole:
\begin{eqnarray}
G^\Delta_{M1}(Q^2) &=& G^\Delta_{M1}(0) \, e^{- Q^2 / \Lambda_{M1}^2} , 
\nonumber \\
G^\Delta_{E2}(Q^2) &=& G^\Delta_{E2}(0) \, e^{- Q^2 / \Lambda_{E2}^2}.  
\label{eq:gm1ge2lattice} 
\end{eqnarray}
The fit to the lattice calculations found as values~\cite{Alexandrou:2009hs}:
\bea
G^\Delta_{M1}(0) &=& 3.04 \pm 0.24, \quad  \Lambda_{M1}^2 = 0.935 \pm 0.122~\mathrm{GeV}^2, \nonumber \\   
G^\Delta_{E2}(0) &=& -2.06^{+1.27}_{-2.35},  \quad  \Lambda_{E2}^2 =   0.54^{+1.69}_{-0.25}~\mathrm{GeV}^2.
\nonumber \\
\eea
The magnetic octupole form factor $G^\Delta_{M3}$ was found to be 
compatible with zero within the statistical accuracy obtained in Ref.~\cite{Alexandrou:2009hs}, and will be neglected in our calculation.

\subsection{$S_{11}(1535)$ intermediate state contribution}

In this section we consider the contribution to $B_n$ when the intermediate state corresponds with the $S_{11}(1535)$ resonance.  The $S_{11}(1535)$ resonance, with mass $M_S = 1.535$~GeV, and quantum numbers $I = 1/2$ and $J^P = 1/2^-$, is the negative parity partner 
of the nucleon. 

A Lorentz-covariant decomposition of the
matrix element of the e.m. current operator $J^\mu$ for the 
$\gamma^\ast N S_{11}$ transition, satisfying manifest e.m. 
gauge-invariance, can be written as:
\begin{eqnarray}
&&\langle S_{11}(p_S, \lambda_S) | J^\mu(0) | N(p, \lambda) \rangle \nn \\
&&\equiv 
\bar \psi (p_S, \lambda_S) 
\Gamma^{\mu}_{N S}(p_S, p)
u(p,\lambda) , 
\eea
where $\psi$ is the spinor for the $S_{11}$ field,
$p_S$ ($\lambda_S$) its four-momentum (helicity) respectively, 
and where the vertex $\Gamma^{\mu}_{N S}$ is given by:
\begin{eqnarray}
\Gamma^{\mu}_{N S}(p_S, p) &=&  
F_1^{N S}
\left( \gamma^\mu - \gamma \cdot q \, \frac{q^\mu}{q^2} \right) \gamma_5 \nonumber \\
&+& F_2^{N S} \frac{i \sigma^{\mu\nu} q_\nu}{(M_N + M_S)} \, \gamma_5  ,
\label{eq:nnstartree}
\end{eqnarray}
with $q \equiv p_S-p$.
The functions $F^{N S}_{1,2}$ are the e.m. FFs for the $\gamma^\ast N S_{11}$ transition and depend on $Q^2$.

Equivalently, one can parametrize the $\gamma^\ast N S_{11}$
transition through two helicity amplitudes $A_{1/2}$ and
$S_{1/2}$, which are defined in the $S_{11}$ rest frame.
These $S_{11}$ rest frame helicity amplitudes are defined
through the following matrix elements of the e.m. current
operator:
\begin{eqnarray}
A_{1/2}^{N S} &\equiv& N_{NS} \langle  S_{11}(\vec 0,  +1/2)
\,|\, J_\mu \cdot \epsilon^\mu_{\lambda = +1} \,|\, N(-\vec q,  -1/2 ) \rangle,  \nonumber \\
S_{1/2}^{N S} &\equiv& N_{NS} \langle  S_{11}(\vec 0,  +1/2)
\,|\, J^0 \,|\, N(-\vec q,  +1/2 ) \rangle,
\label{eq:s11resthel}
\end{eqnarray}
where both spinors are chosen to have the indicated spin projections 
along the $z$-axis (which is chosen along the
virtual photon direction) and where the transverse photon
polarization vector entering $A_{1/2}$ is given by ${\vec 
\epsilon}_{\lambda = +1} = -1/\sqrt{2} (1, i, 0)$. Furthermore in
Eq.~(\ref{eq:s11resthel}), we introduced the conventional normalization factor
\begin{equation}
N_{NS} \equiv  \frac{e}{\sqrt{4 M_N (M_S^2 - M_N^2)}} .
\label{eq:NNS}
\end{equation}

The helicity amplitudes are also functions of the photon virtuality $Q^2$ and 
have been extracted from data on the pion electroproduction process on the proton. 
Using the empirical parameterizations of the 
helicity amplitudes $A_{1/2}^{pS}$, and $S_{1/2}^{pS}$ from Ref.~\cite{Tiator:2011pw}, which are listed in Eq.~(\ref{eq:helmaid}), 
the transition FFs can then be obtained as: 
\begin{eqnarray}
&&F _1^{N S} = \frac{Q^2}{\sqrt{2} N_{NS} Q_{NS+}  Q_{NS-}^2} \nonumber \\
&&\hspace{.35cm} \times \left\{ A_{1/2}^{N S}  - (M_S - M_N) \sqrt{2} \left( \frac{2 M_S}{Q_{NS+} Q_{NS-}} \right) S_{1/2}^{N S} \right\} , 
\nonumber \\
&&F_2^{N S} = \frac{(M_S^2 - M_N^2)}{\sqrt{2} N_{NS} Q_{NS+} Q_{NS-}^2}  \nonumber \\
&&\hspace{.35cm} \times 
\left\{ A_{1/2}^{N S}  + \frac{Q^2}{ (M_S - M_N)} \sqrt{2} \left( \frac{2 M_S}{Q_{NS+}  Q_{NS-} } \right) S_{1/2}^{N S} \right\} ,
\nonumber \\
\end{eqnarray}
where we generalized the shorthand notation of Eq.~(\ref{eq:qpm}) as: 
\bea
Q_{ij\pm}^2 \equiv Q^2 + (M_i \pm M_j)^2 ,
\eea 
with $i, j = N, \Delta, S, D$ denoting the $N$, $\Delta$, $S_{11}$, $D_{13}$ states in the following.  

A Lorentz-covariant decomposition of the
matrix element of the e.m. current operator $J^\mu$ for
the transition $\gamma^\ast S_{11}\Delta$, satisfying manifest e.m. 
gauge-invariance, can be written as:
\begin{eqnarray}
&&\langle S_{11} (p_S, \lambda_S) | J^\mu(0) | \Delta (p_\Delta, \lambda_\Delta) 
\rangle  \nn \\
&&\equiv \bar \psi (p_S,\lambda_S) 
\Gamma^{\alpha \mu}_{\Delta S}(p_S, p_\Delta)
u_\alpha(p_\Delta,\lambda_\Delta) , 
\eea
where the vertex $\Gamma^{\alpha \mu}_{\Delta S}$ is given 
by:  
\begin{eqnarray}
\Gamma^{\alpha \mu} _{ \Delta S}(p_S,p_\Delta)
&=&\, \frac{1}{Q_{\Delta S-} Q_{\Delta S+}} \nn \\
&\times& \left\{ 
\, \left(q^\alpha \,  \gamma^\mu -\, \gamma \cdot q \, g^{\alpha \mu} \right) M_\Delta
\, F_1^{\Delta S} \right. \nonumber \\
&& +\left(
q^\alpha \, P^\mu - q \cdot P \, g^{\alpha \mu}  \right) 
\,  F_2^{\Delta S} 
\nonumber \\
&&\left. 
+ \left(q^\alpha \, q^\mu - q^2 \, g^{\alpha \mu} \right) 
F_3^{\Delta S}  \right\}, \quad
\label{eq:S11del}
\end{eqnarray}
where $P \equiv (p_\Delta + p_S)/2$ and $q \equiv p_S - p_\Delta$. 
In the definition of Eq.~(\ref{eq:S11del}), the FFs are defined for the $\Delta^+ \to S_{11}$ transition, 
and the prefactor $1/ (Q_{\Delta S-} Q_{\Delta S+})$ was chosen such that the 
resulting e.m. FFs $F_{1,2,3}^{\Delta S}$  are 
dimensionless.

The helicity amplitudes are defined through the following specific matrix elements of the electromagnetic current operator
\begin{align}
A_{-1/2}^{\Delta S} &\equiv N_{\Delta S} \langle  S_{11}(\vec 0,  -1/2)
\,|\, J_\mu \cdot \epsilon^\mu_{\lambda = +1} \,|\, \Delta(- \vec q,  -3/2 ) \rangle,  \nonumber \\
A_{1/2}^{\Delta S} &\equiv N_{\Delta S} \langle \; S_{11}(\vec 0,  +1/2)
\,|\, J_\mu \cdot \epsilon^\mu_{\lambda = +1} \,|\, \Delta(- \vec q,  -1/2 ) \rangle,  \nonumber \\
S_{1/2}^{\Delta S} &\equiv N_{\Delta S} \langle \; S_{11}(\vec 0,  +1/2)
\,|\, J^0 \,|\, \Delta(- \vec q,  +1/2 ) \rangle,
\label{eq:SDhelicity}
\end{align}
where the subscripts on the helicity amplitudes indicate the $S_{11}$ spin projections 
along the $z$-axis (which is chosen along the virtual photon direction), and  
where we introduced the normalization factor
\begin{equation}
N_{\Delta S} \equiv  \frac{e}{\sqrt{4 M_S (M_S^2 - M_\Delta^2)}} .
\label{eq:NSDel}
\end{equation}
Note that we can relate the above helicity amplitudes $A^{\Delta B}_{\lambda_B}$ for the $\Delta \to B$ transition, in the rest frame of the baryon resonance $B$ with helicity $\lambda_B$, to the corresponding amplitudes $A^{B \Delta}_{\lambda_\Delta}$  for the $B \to \Delta$ transition, in the rest frame of the $\Delta$ with  helicity $\lambda_\Delta$, as:
\bea 
A^{B \Delta}_{\lambda_\Delta} = \eta_{B} \eta_{\Delta} A^{\Delta B}_{1 - \lambda_\Delta}  ,
\eea
with $\eta_B, \eta_\Delta$ the corresponding intrinsic parities. 

The relations between the helicity amplitudes of Eq.~(\ref{eq:SDhelicity}) and the transition FFs for the electromagnetic 
$\Delta \to S_{11}$ transition can be obtained as:
\begin{widetext}
\begin{align}
\label{eq:SDrelate}
F_1^{\Delta S}  &= - \frac{1}{N_{\Delta S} \, Q_{\Delta S -} }
\left[  \sqrt{3} A_{1/2}^{\Delta S}  -  A_{-1/2}^{\Delta S}  \right], \nonumber \\
F_2^{\Delta S}  &=   \frac{1 }{N_{\Delta S} \, Q_{ \Delta S-} }
\left[  \sqrt{3} A_{1/2}^{\Delta S}  -  A_{-1/2}^{\Delta S}  \right] 
- \frac{ (Q^2 + M_S^2 - M_\Delta^2 ) }{N_{\Delta S} Q_{\Delta S+}^2  Q_{\Delta S-}}  \left[ \sqrt{3} A_{1/2}^{\Delta S} +  A_{-1/2}^{\Delta S} \right]  
-  \frac{4 \sqrt{6} M_\Delta M_S Q^2}{N_{\Delta S} Q_{\Delta S+}^3 Q_{\Delta S-}^2 } S_{1/2}^{\Delta S}, \nonumber \\
F_3^{\Delta S}  &= - \frac{1}{2 N_{\Delta S} \, Q_{ \Delta S-} }
\left[  \sqrt{3} A_{1/2}^{\Delta S}  -  A_{-1/2}^{\Delta S}  \right] 
+ \frac{ (Q^2 + M_S^2 + 3 M_\Delta^2 ) }{2 N_{\Delta S} Q_{\Delta S+}^2  Q_{ \Delta S-}}  \left[ \sqrt{3} A_{1/2}^{\Delta S} +  A_{-1/2}^{\Delta S} \right]  
-  \frac{2 \sqrt{6} M_\Delta M_S (M_S^2 - M_\Delta^2)}{N_{\Delta S} Q_{\Delta S+}^3 Q_{\Delta S-}^2 } S_{1/2}^{\Delta S}.
\end{align}
\end{widetext}

As the helicity amplitudes $A_{-1/2}^{\Delta S}$, $A_{1/2}^{\Delta S}$, and $S_{1/2}^{\Delta S}$ are not know from experiment, we will estimate them using a non-relativistic quark model, as detailed in the Appendix. The quark model provides relations between the helicity amplitudes for the $\Delta \to S_{11}$ transition and the corresponding ones for the $p \to S_{11}$ and $p \to D_{13}$ transitions, as given by Eqs.~(\ref{eq:qmrel1},\ref{eq:qmrel2}).  For the numerical estimates, we will use these relations and use the empirical results of Eq.~(\ref{eq:helmaid}) for the 
electromagnetic $p \to S_{11}$ and $p \to D_{13}$ helicity amplitudes as input. 

The inelastic contribution to $\hat W^{\beta \mu \nu}$,
corresponding with the $S_{11}(1535)$   
intermediate state,
can then be expressed in terms of on-shell $\gamma^\ast N S_{11}$ and $\gamma^\ast \Delta S_{11} $ 
vertices as:
\bea
\hat{W}^{\beta \mu \nu}_{S_{11}}(p^\prime,p) 
&=& 2\pi \, \delta(W^2-M_S^2) \;  \tilde \Gamma^{\beta\mu}_{\Delta S}(p_S, p^\prime)   \nonumber \\
& \times &  P^{(1/2)}(p_S, M_S)  \Gamma^{\nu}_{N S}(p_S, p)  , \quad \quad
\label{eq:wS11}
\eea
where the adjoint vertex $\tilde \Gamma^{\beta \mu}_{\Delta S} \equiv \gamma^0 
\left(\Gamma^{ \beta \mu}_{\Delta S}\right)^\dagger \gamma^0$ is given by exactly the same 
operator as in Eq.~(\ref{eq:S11del}), with $q = p_S - p^\prime$ in this case denoting the outgoing photon momentum. 

This allows us, for the $S_{11}$ intermediate state contribution, to evaluate the unpolarized hadronic tensor 
entering Eqs.~(\ref{eq:an2}, \ref{eq:an3}) for $B_n$ as:
\bea
H^{\kappa\mu\nu}_{S_{11}} &=&  2\pi \, \delta(W^2-M_S^2) \, \nonumber \\
&\times& \mathrm{Tr} \left\{\tilde \Gamma^{\alpha\kappa}_{N \Delta}(p^\prime, p)  P^{(3/2)}_{\alpha \beta}(p^\prime,M_\Delta) 
\tilde \Gamma^{\beta \mu}_{\Delta S}(p_S, p^\prime) \right. \nonumber \\ 
& \times& \left. P^{(1/2)}(p_S,M_S) 
 \Gamma^{\nu}_{NS}(p_S, p) P^{(1/2)}(p,M_N) \right\} . \;\;\;\;\;\;
\eea

\subsection{$D_{13}(1520)$ intermediate state contribution}

We next consider the contribution to $B_n$ when the intermediate state corresponds with the $D_{13}(1520)$ resonance.  This is  
the lowest mass baryon resonance, with mass $M_D = 1.520$~GeV, 
which has quantum numbers $I = 1/2$ and $J^P = 3/2^-$.

A Lorentz-covariant decomposition of the
matrix element of the e.m. current operator $J^\mu$ for
the $\gamma^\ast N D_{13}$ transition, satisfying manifest e.m. 
gauge-invariance, is given by:
\begin{eqnarray}
&&\langle D_{13}(p_D, \lambda_D) | J^\mu(0) | N (p, \lambda) 
\rangle \nn \\
&& \equiv 
\bar \psi_\alpha (p_D,\lambda_D) 
\Gamma^{\alpha  \mu}_{N D}(p_D, p)
u(p,\lambda),
\eea
with $p_D$ ($\lambda_D$) denoting the four-momentum (helicity) of the $D_{13}$ state respectively, 
where $\psi_\alpha$ is the Rarita-Schwinger spinor for the $D_{13}$ field, and where 
the vertex $\Gamma^{\alpha  \mu}_{N D}$ is given by:
\begin{eqnarray}
\Gamma^{\alpha  \mu}_{ND}(p_D, p) &=&
\frac{1}{Q_{ND -} Q_{ND +}} \nn \\
&\times& \left\{
\left( q^\alpha \, \gamma^{\mu} -q \cdot \gamma   g^{\alpha \mu} \right) M_D 
F_1^{N D}  \right. 
\nonumber \\
&&  +  \,
\left( q^\alpha \, {p_D}^{\mu} -q \cdot p_D  \, g^{\alpha \mu} \right) 
F_2^{N D}  
\nonumber \\
&& \left.  + \,
\left( q^\alpha \, q^\mu - q^2 \, g^{\alpha \mu}  \right) 
F_3^{N D}  
\right\} , 
\label{eq:d13ff}
\end{eqnarray}
with $q \equiv p_D - p$. 
In Eq.~(\ref{eq:d13ff}), the prefactor was chosen such that the 
resulting e.m. FFs $F_{1,2,3}^{ND}$  are 
dimensionless. 

In the same way as we did for the $\gamma^\ast N S_{11}$ transition above, 
one can also parametrize the $\gamma^\ast N D_{13}$
transition through helicity amplitudes in the $D_{13}$ rest frame. 
For the spin-3/2 resonance, we need three helicity amplitudes $A_{3/2}^{ND}$, $A_{1/2}^{ND}$ and
$S_{1/2}^{ND}$, which are defined
through the following matrix elements of the e.m. current operator:
\begin{eqnarray}
A_{3/2}^{ND} &\equiv& N_{ND} \langle  D_{13}(\vec 0, +3/2)
|\,  J_\mu \cdot \epsilon^\mu_{\lambda = +1} \,| N(-\vec q, +1/2 ) \rangle,  \nonumber \\
A_{1/2}^{ND} &\equiv& N_{ND} \langle \; D_{13}(\vec 0, +1/2)
|\,  J_\mu \cdot \epsilon^\mu_{\lambda = +1} \,| N(-\vec q, -1/2 ) \rangle,  \nonumber \\
S_{1/2}^{ND} &\equiv& N_{ND} \langle \; D_{13}(\vec 0, +1/2)
|\, J^0 \,| N(-\vec q, +1/2 ) \rangle,
\label{eq:d13resthel}
\end{eqnarray}
with $N_{ND}$ defined, analogously as in Eq.~(\ref{eq:NNS}), as
\begin{equation}
N_{ND} \equiv  \frac{e}{\sqrt{4 M_N (M_D^2 - M_N^2)}} .
\label{eq:NND}
\end{equation}
Using the empirical parameterizations of the 
helicity amplitudes $A_{3/2}^{pD}$, $A_{1/2}^{pD}$, and $S_{1/2}^{pD}$ from Ref.~\cite{Tiator:2011pw}, 
which are listed in Eq.~(\ref{eq:helmaid}), 
the transition FFs can then be obtained as: 
\begin{eqnarray}
F_1^{N D}  &=& \frac{ 1}{N_{ND} Q_{ND -}} 
\left\{  A_{3/2}^{ND} - \sqrt{3} A_{1/2}^{ND}  \right\}, \nonumber \\
F_1^{N D}  + F_2^{N D}  &=& \frac{1}{N_{ND} Q_{ND +}^2  Q_{ND -} }\nonumber \\
& \times&
\left\{ (M_D^2 - M_N^2 - Q^2)  \left[ A_{3/2}^{ND} + \sqrt{3} A_{1/2}^{ND} \right]  \right. \nonumber \\
&&\left. \hspace{.2cm} -  \frac{4 \sqrt{6} M_D^2 Q^2}{Q_{ND+} Q_{ND -}} S_{1/2}^{ND} \right\}, \nonumber \\
F_3^{N D}  &=& - \frac{2 M_D^2}{N_{ND} Q_{ND+}^2 Q_{ND-}} 
\left\{ A_{3/2}^{ND} + \sqrt{3} A_{1/2}^{ND}  \right. \nonumber \\
&+&\left.   \frac{\sqrt{6} (M_D^2 - M_N^2 - Q^2)}{Q_{ND+} Q_{ND-} }  S_{1/2}^{ND} \right\}. \quad
\end{eqnarray}

A Lorentz-covariant decomposition for the 
on-shell $\gamma^* \Delta D_{13}$ vertex which exhibits
manifest electromagnetic gauge-invariance as: 
\begin{eqnarray}
&&\langle D_{13} (p_D, \lambda_D) | J^\mu(0) | \Delta (p_\Delta, \lambda_\Delta) 
\rangle  \nn \\
&&\equiv \bar \psi_\beta (p_D,\lambda_D) 
\Gamma^{\alpha \beta \mu}_{\Delta D}(p_D, p_\Delta)
u_\alpha(p_\Delta,\lambda_\Delta) , 
\eea
where $p_\Delta$ ($p_D$) are the four-momenta and $\lambda_\Delta$ ($\lambda_{D}$) the helicities of $\Delta$ ($D_{13}$) respectively, and 
where the vertex $\Gamma^{\alpha \beta \mu}_{\Delta D}$ is given 
by:  
\bea 
&&\Gamma^{\alpha \beta \mu}_{\Delta D}(p_D, p_\Delta) = \nonumber \\
&&- \left[  F_1^{\Delta D}  g^{\alpha \beta}  
+ F_3^{\Delta D} \frac{q^\alpha q^\beta}{(M_\Delta+M_D)^2} 
\right] \left( \gamma^\mu - \gamma \cdot q \, \frac{q^\mu}{q^2} \right)  \gamma_5 \nonumber \\
&&- \left[ F_2^{\Delta D}  g^{\alpha \beta}
+ F_4^{\Delta D} \frac{q^\alpha q^\beta}{(M_\Delta+M_D)^2}\right] 
\frac{i \sigma^{\mu\nu} q_\nu}{(M_\Delta + M_D)}\, \gamma_5 \nonumber \\
&&- \frac{F_5^{\Delta D}}{(M_\Delta + M_D)} \left( g^{\alpha \mu} q^\beta - g^{\beta \mu} q^\alpha \right) \gamma_5,
\label{eq:gaDdeltree}
\end{eqnarray}
where $q \equiv p_D - p_\Delta$. 

Although we will only need on-shell vertices in this work, one can also define consistent vertices for off-shell spin-3/2 particles which satisfy a spin-3/2 gauge invariance, as discussed in Ref.~\cite{Pascalutsa:1998pw,Pascalutsa:1999zz}, i.e. $(p_{\Delta)_\alpha} \Gamma^{\alpha \beta \mu}_{\Delta D} = 0$ and 
$(p_D)_\beta \Gamma^{\alpha \beta \mu}_{\Delta D} = 0$, by replacing \textit{e.g.} in Eq.~(\ref{eq:gaDdeltree}):
\begin{eqnarray}
g^{\alpha \beta} &\to& \frac{1}{M_\Delta^2 M_{D}^2} \left\{ p_\Delta^{2} p_D^2 g^{\alpha \beta} - p_\Delta^{2} p_D^\alpha p_D^\beta 
- p_D^2 p_\Delta^{\alpha} p_\Delta^\beta \right. \nonumber \\
&&\left. \hspace{1.5cm} + p_\Delta \cdot p_D p_\Delta^\alpha p_D^\beta \right\},
\end{eqnarray}
or
\begin{align}
q^\alpha q^\beta \to \left( q^\alpha - \frac{ q \cdot p_\Delta }{ p_\Delta^2 } p_\Delta^\alpha \right)
		\left( q^\beta - \frac{ q \cdot p_D }{ p_D^2 } p_D^\beta \right)	.
\end{align}

For the $\Delta \to D_{13}$ amplitude, there are five helicity amplitudes, defined by the following matrix elements of the e.m. current operator,
\begin{align}
A_{3/2}^{\Delta D} &\equiv  N_{\Delta D} \langle  D_{13}(\vec 0, +3/2)
|  J_\mu \cdot \epsilon^\mu_{\lambda = +1} \,|\, \Delta(- \vec q, +1/2 ) \rangle,  \nonumber \\
A_{1/2}^{\Delta D} &\equiv  N_{\Delta D} \langle D_{13}(\vec 0, +1/2)
\,|\,  J_\mu \cdot \epsilon^\mu_{\lambda = +1} \,|\, \Delta(- \vec q, -1/2 ) \rangle,  \nonumber \\
A_{-1/2}^{\Delta D} &\equiv N_{\Delta D} \langle D_{13}(\vec 0,  -1/2)
\,|\,  J_\mu \cdot \epsilon^\mu_{\lambda = +1} \,|\, \Delta(- \vec q,  -3/2 ) \rangle,  \nonumber \\
S_{3/2}^{\Delta D} &\equiv N_{\Delta D} \langle  D_{13}(\vec 0,  +3/2)
\,|\, J^0 \,|\, \Delta(- \vec q,  +3/2 ) \rangle,		\nn\\
S_{1/2}^{\Delta D} &\equiv N_{\Delta D} \langle  D_{13}(\vec 0,  +1/2)
\,|\, J^0 \,|\, \Delta(- \vec q,  +1/2 ) \rangle,
\label{eq:DDhelicity}
\end{align}
where $N_{\Delta D}$ is defined as
\begin{align}
N_{\Delta D} \equiv \frac{ e }{ \sqrt{ 4 M_D ( M_D^2 - M_\Delta^2 ) } }		\,.
\label{eq:NDDel}
\end{align}
It is also convenient to introduce  
\begin{align}
\tilde F_{1,3}^{\Delta D} = F_{1,3}^{\Delta D} + \left( \frac{ M_D - M_\Delta }{ M_D + M_\Delta} \right) F_{2,4}^{\Delta D}.
\end{align}
The helicity amplitudes for the electromagnetic $\Delta \to D_{13}$ transition are obtained as:
\begin{widetext}
\begin{align}
A_{3/2}^{\Delta D} &= N_{\Delta D} \sqrt{ \frac{2}{3} } Q_{\Delta D+}  \left\{ \tilde F_1^{\Delta D} 
	- \frac{ Q_{\Delta D-}^2 }{ 2M_\Delta (M_D + M_\Delta) } F_5^{\Delta D}	\right\} ,	\nn\\
A_{-1/2}^{\Delta D} &= N_{\Delta D} \sqrt{ \frac{2}{3} } Q_{\Delta D+} 
 \left\{ \tilde F_1^{\Delta D} 
	+ \frac{ Q_{\Delta D-}^2 }{ 2M_D (M_D + M_\Delta)} F_5^{\Delta D}	\right\}, \nn \\
A_{1/2}^{\Delta D} &=  N_{\Delta D} \frac{ \sqrt{2} }{6} 	\frac{ Q_{\Delta D+} }{ M_\Delta M_D }	\left\{
	2\left( Q^2 + M_D^2 + M_\Delta^2 \right) \tilde F_1^{\Delta D} 
	- \frac{ Q_{\Delta D+}^2 Q_{\Delta D-}^2 }{ (M_D + M_\Delta )^2 } \tilde F_3^{\Delta D}
	+ \frac{ (M_D - M_\Delta) Q_{\Delta D-}^2 }{(M_D + M_\Delta)} F_5^{\Delta D} 	\right\} ,	\nn\\
S_{3/2}^{\Delta D} &= N_{\Delta D} \frac{Q_{\Delta D+}^2 Q_{\Delta D-} }{ 2 M_D Q^2 }	\left\{
	-   (M_D - M_\Delta ) \tilde F_1^{\Delta D}
	+  \frac{Q_{\Delta D-}^2}{(M_D + M_\Delta)} F_2^{\Delta D}	\right\} ,							\nn\\
S_{1/2}^{\Delta D} &= N_{\Delta D} \frac{ Q_{\Delta D+}^2 Q_{\Delta D-} }{ 6 M_D^2 M_\Delta Q^2 }	\bigg\{
	( Q^2 + M_D^2 + M_\Delta^2 - M_D M_\Delta ) \left[ - (M_D - M_\Delta) \tilde F_1^{\Delta D}
	+ \frac{ Q_{\Delta D-}^2}{(M_D + M_\Delta)} F_2^{\Delta D}	\right] \nn\\
	&	\hskip 8.5 em
	+ \frac{ Q_{\Delta D+}^2 Q_{\Delta D-}^2 }{ 2 (M_D + M_\Delta)^2 } \left[ (M_D - M_\Delta)  \tilde F_3^{\Delta D}
	- \frac{ Q_{\Delta D-}^2 }{ (M_D + M_\Delta)} F_4^{\Delta D} \right] + \frac{ Q_{\Delta D-}^2 Q^2 }{ (M_D + M_\Delta)} F_5^{\Delta D}
	\bigg\} .
	\label{eq:DeltaD13hel}
\end{align}
\end{widetext}


Inverting the relations in Eq.~(\ref{eq:DeltaD13hel}) gives
\begin{widetext}
\begin{align}
\tilde F_1^{\Delta D} &= \sqrt{ \frac{3}{2} }	\frac{ 1 }{ N_{\Delta D} Q_{\Delta D+} (M_D + M_\Delta)}
	\left(	M_D A_{-1/2}^{\Delta D} + M_\Delta A_{3/2}^{\Delta D}  \right)	,		\nn\\
F_2^{\Delta D}	&= \frac{ (M_D + M_\Delta) }{ N_{\Delta D}  Q_{\Delta D+}^2 Q_{\Delta D-}^3 }
	\left\{ 2 M_D Q^2 S_{3/2}^{\Delta D} + \sqrt{ \frac{3}{2} } \, 
	Q_{\Delta D+} Q_{\Delta D-} \frac{( M_D - M_\Delta )}{( M_D + M_\Delta )} 
	\left(M_D  A_{-1/2}^{\Delta D} + M_\Delta A_{3/2}^{\Delta D} \right) \right\}		,\nn\\
\tilde F_3^{\Delta D} &= \sqrt{ 6}
	\frac{ (M_D + M_\Delta) }{ N_{\Delta D} Q_{\Delta D+}^3 Q_{\Delta D-}^2 }
	\Big\{ 	 M_D \left( Q^2 + M_D (M_D + M_\Delta) \right) 	A_{-1/2}^{\Delta D}
+	M_\Delta \left( Q^2 + M_\Delta (M_D + M_\Delta) \right)  A_{3/2}^{\Delta D} 
\nn \\
& \hskip 11em 
- \sqrt{3} M_D M_\Delta (M_D + M_\Delta) A_{1/2}^{\Delta D} 
	\Big\}		,\nn\\
F_4^{\Delta D}	&= 
	\frac{2 (M_D + M_\Delta )^3 }{ N_{\Delta D} Q_{\Delta D+}^4 Q_{\Delta D-}^5 }
	\bigg\{ 2 M_D Q^2 \left[  ( Q^2 + M_D^2 + M_\Delta^2 - M_D M_\Delta)  S_{3/2}^{\Delta D} 
	- 3 M_\Delta M_D  S_{1/2}^{\Delta D} \right] 
				\nn\\
&\hskip 9em + 	\sqrt{ \frac{3}{2} } Q_{\Delta D+} Q_{\Delta D-} \left[   M_D (Q^2 +   M_D ( M_D - M_\Delta ))  A_{- 1/2}^{\Delta D}
-      	M_\Delta (Q^2 -   M_\Delta ( M_D - M_\Delta ))  A_{3/2}^{\Delta D} \right. \nn \\
&\left. \hskip 17em -	\sqrt{ 3 }  M_D M_\Delta ( M_D - M_\Delta )  A_{1/2}^{\Delta D} \right] 
	\bigg\}		,
				\nn\\
F_5^{\Delta D}	&= \sqrt{ 6 }	\frac{ M_D M_\Delta }{ N_{\Delta D} Q_{\Delta D+} Q_{\Delta D -}^2 }
	\left(	A_{-1/2}^{\Delta D} - A_{3/2}^{\Delta D}  \right)		.
\end{align}

\end{widetext}

As discussed above for the electromagnetic $\Delta \to S_{11}$ transition, for our numerical estimates we will 
also use the quark model to relate the helicity amplitudes for the $\Delta \to D_{13}$ transition to the corresponding ones for the $p \to S_{11}$ and $p \to D_{13}$ transitions, as given by Eqs.~(\ref{eq:qmrel1},\ref{eq:qmrel2}), and use the empirical results 
of Eq.~(\ref{eq:helmaid}) for the latter.

The inelastic contribution to $\hat W^{\beta \mu \nu}$,
corresponding with the $D_{13}(1520)$   
intermediate state,
can then be expressed in terms of on-shell $\gamma^\ast N D_{13}$ and $\gamma^\ast \Delta D_{13}$ 
vertices as:
\bea
\hat{W}^{\beta \mu \nu}_{D_{13}}(p^\prime,p) 
&=& 2\pi \, \delta(W^2-M_D^2) \;   \tilde \Gamma^{\beta \gamma \mu}_{\Delta D}(p_D, p^\prime)  \nonumber \\
&\times&   P^{(3/2)}_{\gamma \delta }(p_D, M_D)  \Gamma^{\delta \nu}_{N D}(p_D, p)  , \quad 
\label{eq:wD13}
\eea
where the adjoint vertex $\tilde \Gamma^{\beta \gamma \mu}_{\Delta D} \equiv \gamma^0 
\left(\Gamma^{ \beta \gamma \mu}_{\Delta D}\right)^\dagger \gamma^0$ is given by the same 
operator as in Eq.~(\ref{eq:gaDdeltree}), with $q = p_D - p^\prime$ in this case denoting the outgoing photon momentum, 
and where in addition the sign of the term proportional to the FF $F_5^{\Delta D}$ is reversed. 

This allows us, for the $D_{13}$ intermediate state contribution, to evaluate the unpolarized hadronic tensor 
entering Eqs.~(\ref{eq:an2}, \ref{eq:an3}) for $B_n$ as:
\begin{align}
H^{\kappa\mu\nu}_{D_{13}}&=  2\pi \, \delta(W^2-M_D^2) \, \nonumber \\
&\times \mathrm{Tr} \left\{\tilde \Gamma^{\alpha\kappa}_{N \Delta}(p^\prime, p)  P^{(3/2)}_{\alpha \beta}(p^\prime,M_\Delta) 
\tilde \Gamma^{\beta \gamma \mu}_{\Delta D}(p_D, p^\prime) \right. \nonumber \\ 
& \times \left. P^{(3/2)}_{\gamma \delta }(p_D,M_D) 
 \Gamma^{\delta \nu}_{N D}(p_D, p) P^{(1/2)}(p,M_N) \right\} . \;
\end{align}

\section{Results and discussion}
\label{sec:results}

In this section, we will show estimates for the normal beam SSA using the hadronic model described above, which includes the contributions of $N$, $\Delta(1232)$, $S_{11}(1535)$, and $D_{13}(1520)$ intermediate states. 

To visualize the contributions from different kinematical regions entering Eq.~(\ref{eq:an4}) for $B_n$, 
we will show density plots of the integrand, which are defined through
\bea
B_n\equiv\int \frac{\mathrm{d}Q_{1}^2\mathrm{d}Q_{2}^2}{Q_1^2Q_2^2}I(Q_1^2,Q_2^2). 
\label{eq:bndens}
\eea

\begin{figure}
\includegraphics[width=8cm]{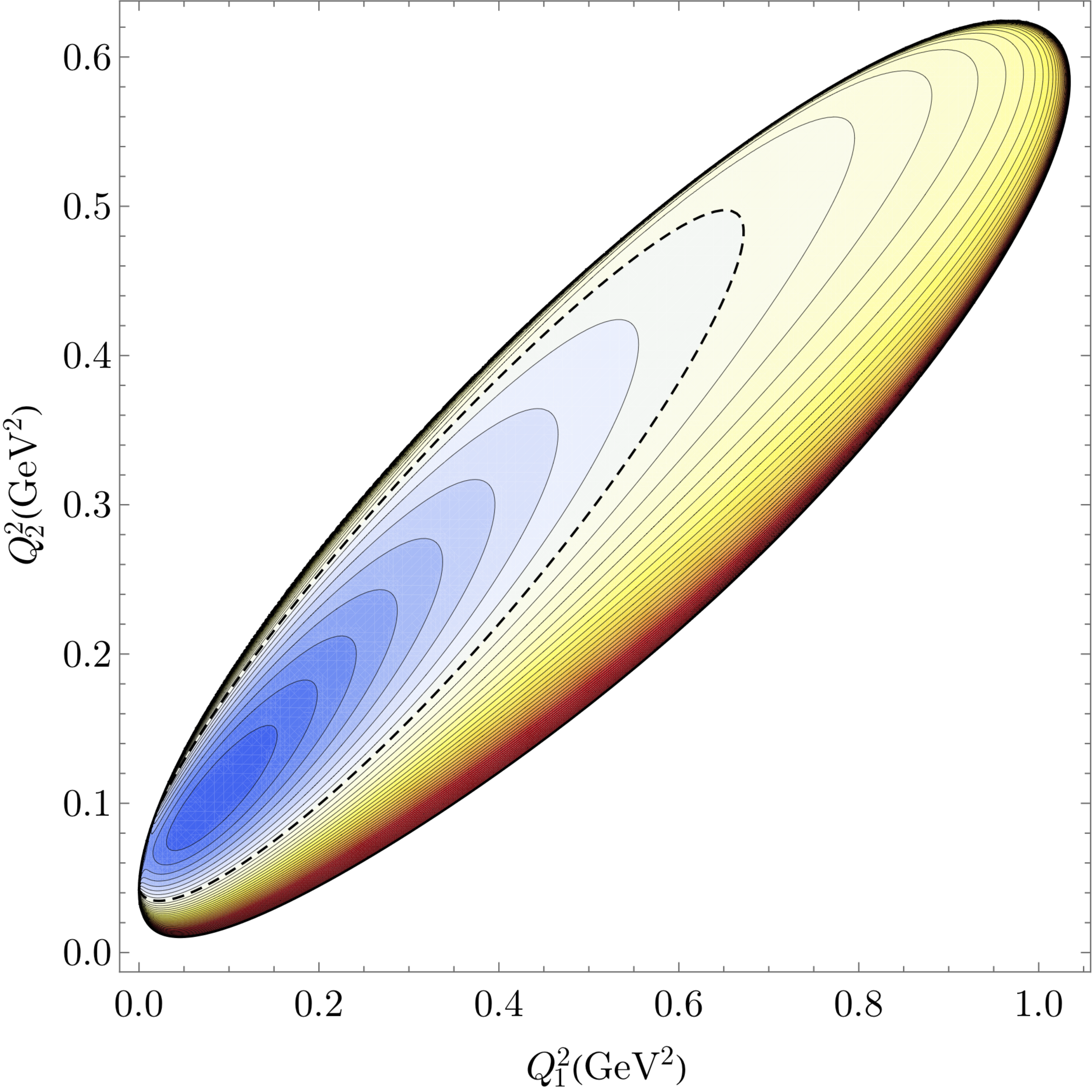}\\
\vspace{0.5cm}
\includegraphics[width=8cm]{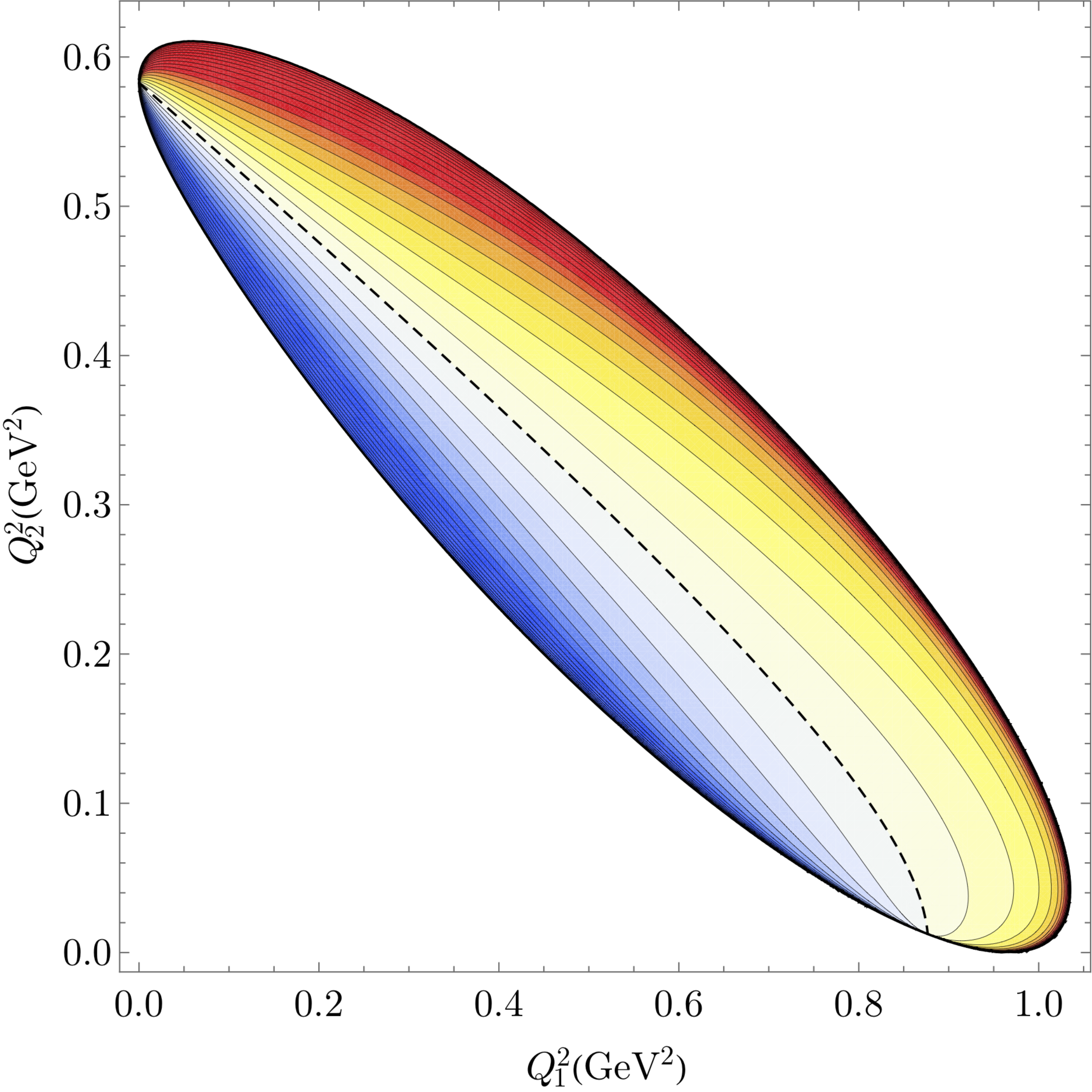}
\caption{Plot of the density $I(Q_1^2,Q_2^2)$ entering the integrand of $B_n$ in Eq.~(\ref{eq:bndens}) for the nucleon intermediate state contribution for $E_e = 0.855$~GeV. The upper and lower panels show the distribution for $\theta_{cm} = 30$~deg and $\theta_{cm} = 150$~deg, respectively. The integrand takes zero value along the dashed curve. Larger negative (positive) values of $I$ correspond with stronger shades of blue (red). The distance between the contours corresponds with $0.5\times 10^{-8}$ for the upper panel and $1.25\times10^{-7}$ for the bottom panel.
}
\label{fig:bn_Q1Q2density_N}
\end{figure}

\begin{figure}
\includegraphics[width=8.cm]{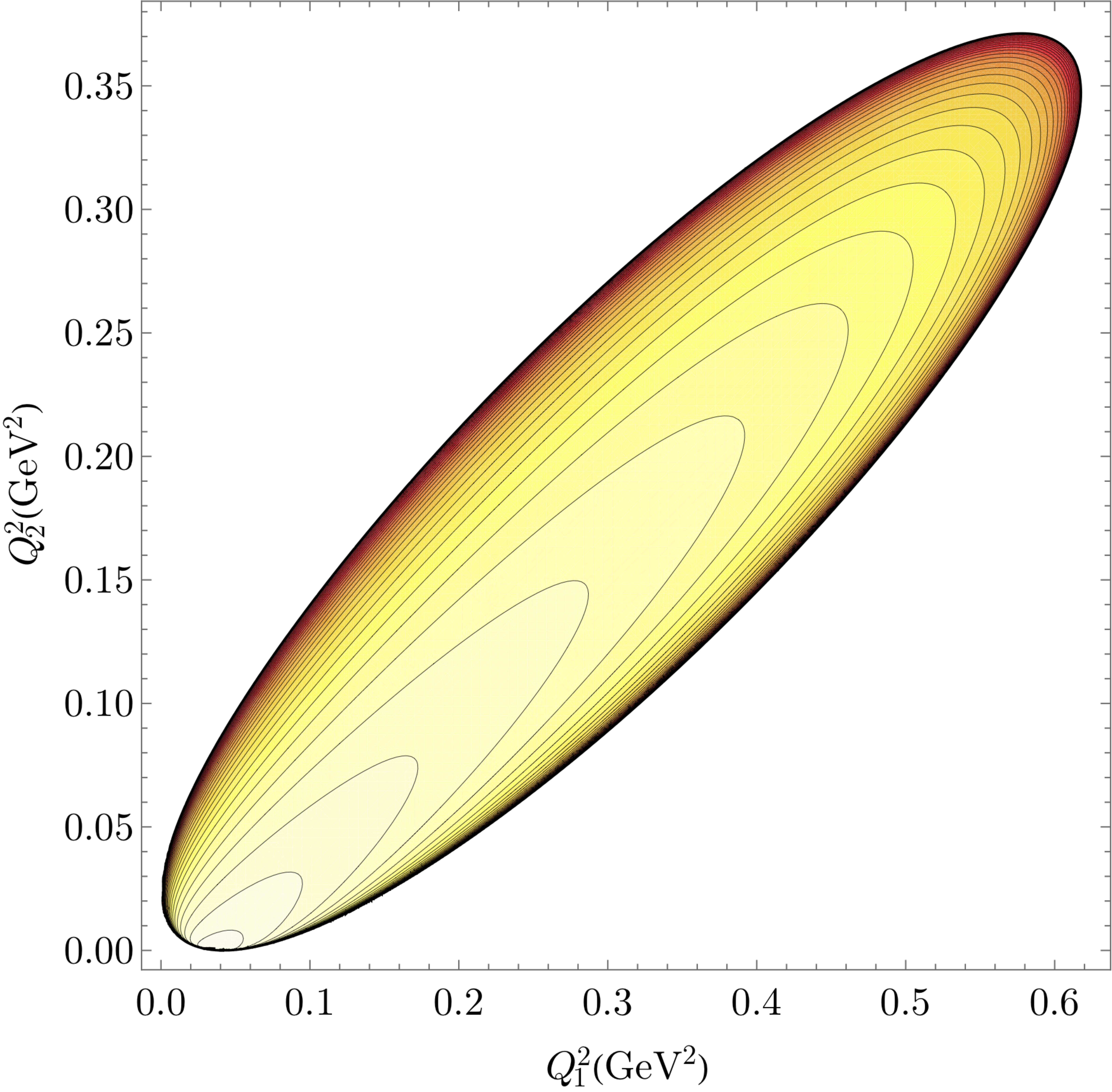}\\
\vspace{0.5cm}
\includegraphics[width=8.cm]{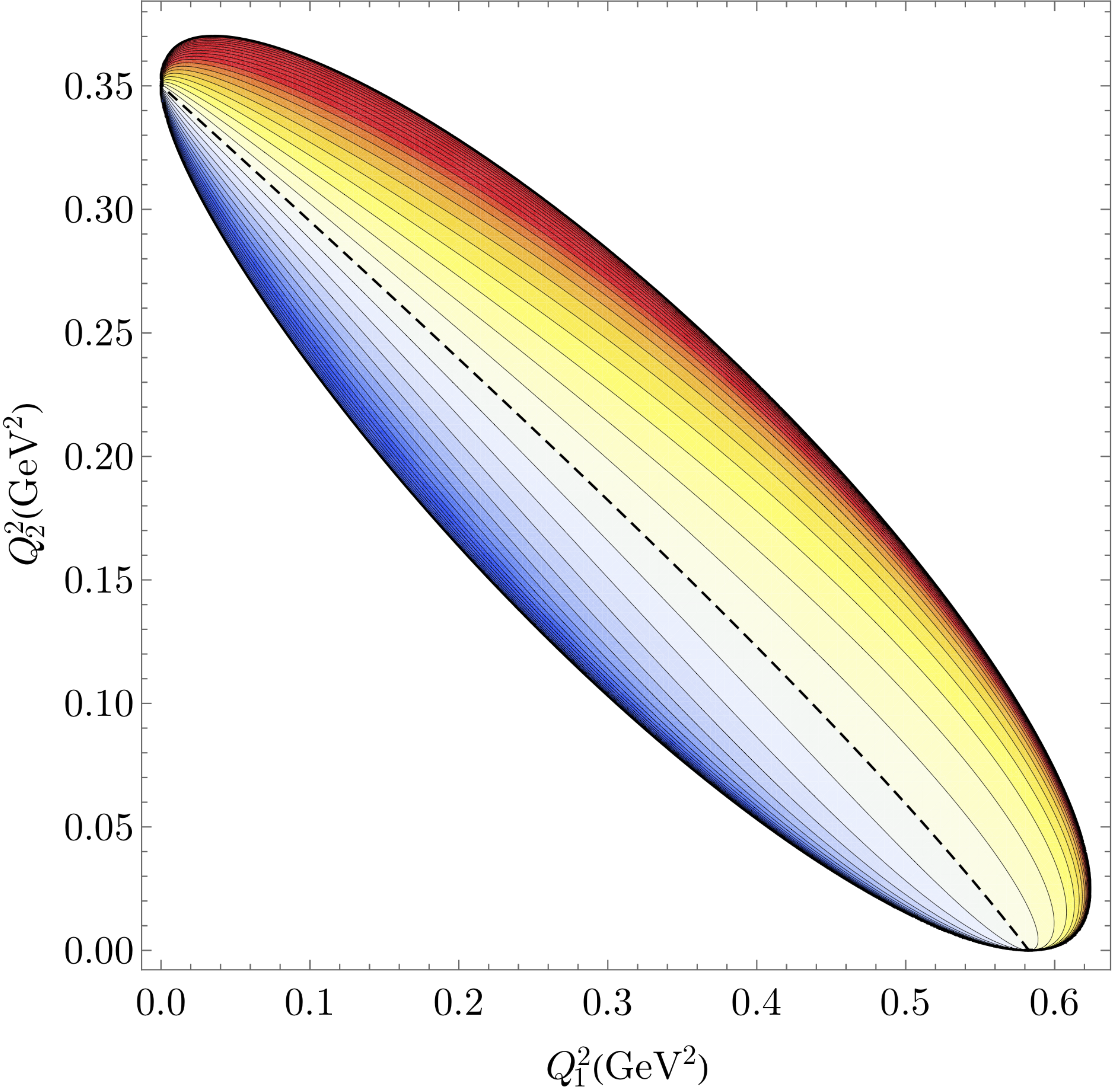}
\caption{Plot of the density $I(Q_1^2,Q_2^2)$ entering the integrand of $B_n$ in Eq.~(\ref{eq:bndens}) for the $\Delta$ intermediate state contribution for $E_e = 0.855$~GeV. The upper and lower panels show the distribution for $\theta_{cm} = 30$~deg and $\theta_{cm} = 150$~deg, respectively. The integrand takes zero value along the dashed curve. Larger negative (positive) values of $I$ correspond with stronger shades of blue (red). The distance between the contours corresponds with $0.5\times 10^{-7}$ for the upper panel and $0.5\times10^{-6}$ for the bottom panel.
}
\label{fig:bn_Q1Q2density_Delta}
\end{figure}

Due to the photon virtualities in the denominator, the full integrand of $B_n$ is very strongly peaked towards the quasi-VCS regions, where either $Q_1^2$ or $Q_2^2$ becomes of order ${\cal O}(m_e^2)$, see Eq.~(\ref{eq:quasivcs}), corresponding with the physical situations where the intermediate electron is collinear with either the incident or scattered electrons.  Furthermore, when $\sqrt s$ approaches the invariant mass $W$ of an intermediate baryon resonance, one also obtains an enhancement as both photons become quasi-real, see Eq.~(\ref{eq:q1real}). 
As the integrand is amplified in the region of small $Q_1^2$ and/or $Q_2^2$ due to these near singularities, special care is needed when integrating over these regions numerically. 

The electromagnetic transition strengths are encoded in the dimensionless density function 
 $I(Q_1^2,Q_2^2)$ in Eq.~(\ref{eq:bndens}). Using the model for the hadronic tensor outlined in Section V, we show the density functions $I(Q_1^2, Q_2^2)$ for a beam energy $E_e = 0.855$~GeV of the A4@MAMI experiment,  in Figs.~\ref{fig:bn_Q1Q2density_N} and \ref{fig:bn_Q1Q2density_Delta} for the $N$ and $\Delta(1232)$  
 intermediate states respectively.

In Fig.~\ref{fig:qweak}, we show our result for the angular dependence of $B_n$ for a beam energy $E_e = 1.165$~GeV, 
corresponding with the Qweak@JLab experiment~\cite{Nuruzzaman:2015vba}.
We notice from Fig.~\ref{fig:qweak} that the nucleon and $\Delta$ intermediate state contributions to $B_n$ are strongly forward peaked. This behavior for the $e p \to e \Delta$ process is unlike the corresponding $B_n$ for the elastic process. 
The measured value for $B_n$ for the elastic $ep \to ep$ process ranges from a few ppm in the forward angular range to around a hundred ppm in the backward angular range for beam energies below and around 1 GeV~\cite{Wells:2000rx, Maas:2004pd, Rios:2017vsw, Armstrong:2007vm, Androic:2011rh, Abrahamyan:2012cg, Waidyawansa:2013yva}, in good  agreement with theoretical TPE expectations~\cite{Pasquini:2004pv}. 
For the inelastic process $e p \to e \Delta$, we expect an enhancement of $B_n$ in the forward angular range, corresponding with  low $Q^2$, since the 
OPE process which enters the denominator of $B_n$, is suppressed by one power of $Q^2$ relative to its elastic counterpart, as seen from Eqs.~(\ref{eq:d1gainel}, \ref{eq:d1gael}).
We furthermore see from Fig.~\ref{fig:qweak} that the sum of $S_{11}(1535) +  D_{13}(1520)$ contributions do not show such forward angular enhancement as their electromagnetic transitions are suppressed by an extra momentum transfer.  The $S_{11}$ and $D_{13}$ contributions show a similar size and strength, and their combined contribution to $B_n$  becomes larger than the $\Delta(1232)$ contribution for angles $\theta_{\rm lab} > 45$~deg.  

In Fig.~\ref{fig:qweak}, we also show a first data point for the beam normal SSA for the $e^- p \to e^- \Delta^+(1232)$ process which has been reported by the Qweak Coll.~\cite{Nuruzzaman:2015vba}. Despite its large error bar, the data point at a forward angle of $\theta_{lab} = 8.3$~deg shows a large value of $B_n$ of around 40~ppm for this process. The data point is very well described both in sign and magnitude by our calculation, confirming the large expected enhancement in the forward angular range. Since the 
$S_{11}(1535) +  D_{13}(1520)$ contribution is very small at this angle, $B_n$ is dominated by $N$ and $\Delta$ intermediate states at this forward angle. Furthermore, since the $N \to N$, $N \to \Delta$ electromagnetic transitions are well known from experiment, and the $\Delta \to \Delta$ electromagnetic transition is completely dominated by the coupling to the $\Delta^+$ charge at this forward angle, the model dependence in our prediction is very small at this angle.

\begin{figure}
\includegraphics[width=9.cm]{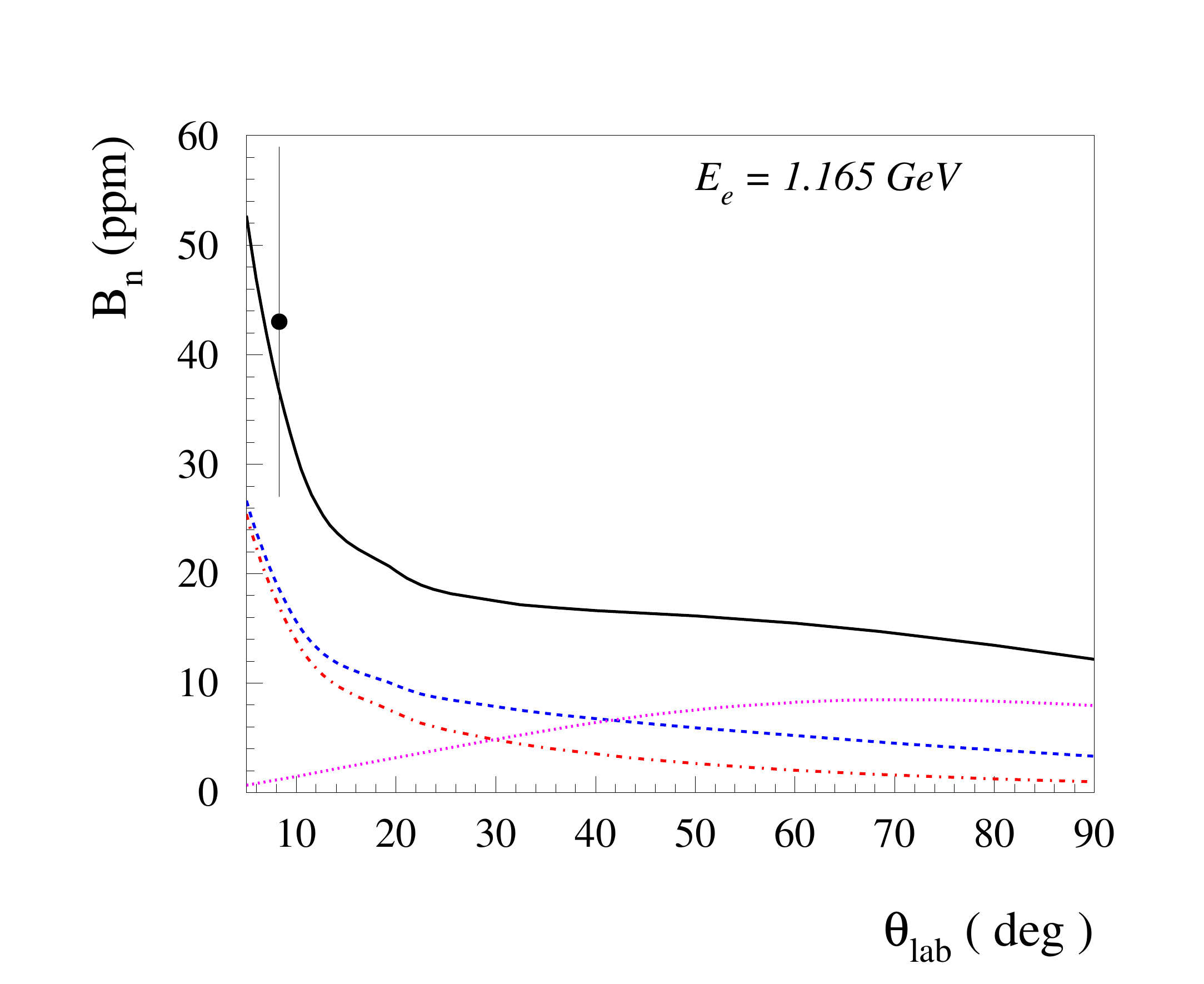}
\caption{Beam normal spin asymmetry $B_n$ for the $e^- p \to e^- \Delta^+$ process 
as function of the {\it lab} scattering angle for a beam energy $E_e = 1.165$~GeV. 
The curves denote the contributions from different intermediate states: 
nucleon (dashed-dotted red curve); 
$\Delta(1232)$ (dashed blue curve);
$S_{11}(1535) + D_{13}(1520)$ (dotted violet curve); 
$N + \Delta + S_{11}(1535) + D_{13}(1520)$ (solid black curve). 
The data point is from the Qweak Coll.~\cite{Nuruzzaman:2015vba}. 
}
\label{fig:qweak}
\end{figure}

In Figs.~\ref{fig:a4mami1} and \ref{fig:a4mami2}, we show the corresponding results for different kinematics corresponding with the 
A4@MAMI experiment. Fig.~\ref{fig:a4mami1} shows the result for $E_e = 0.855$~GeV. This beam energy corresponds with a value $\sqrt s \approx 1.58$~GeV, which is closer to the $S_{11}(1535)$ and $D_{13}(1520)$ thresholds. We therefore expect an enhancement of their contributions. As one gets very close to the threshold for an intermediate state contribution, one approaches the situation where the intermediate electron becomes soft, and both photons have small virtualities, see Eq.~(\ref{eq:q1real}), corresponding with the quasi-real Compton process.

\begin{figure}
\includegraphics[width=9.cm]{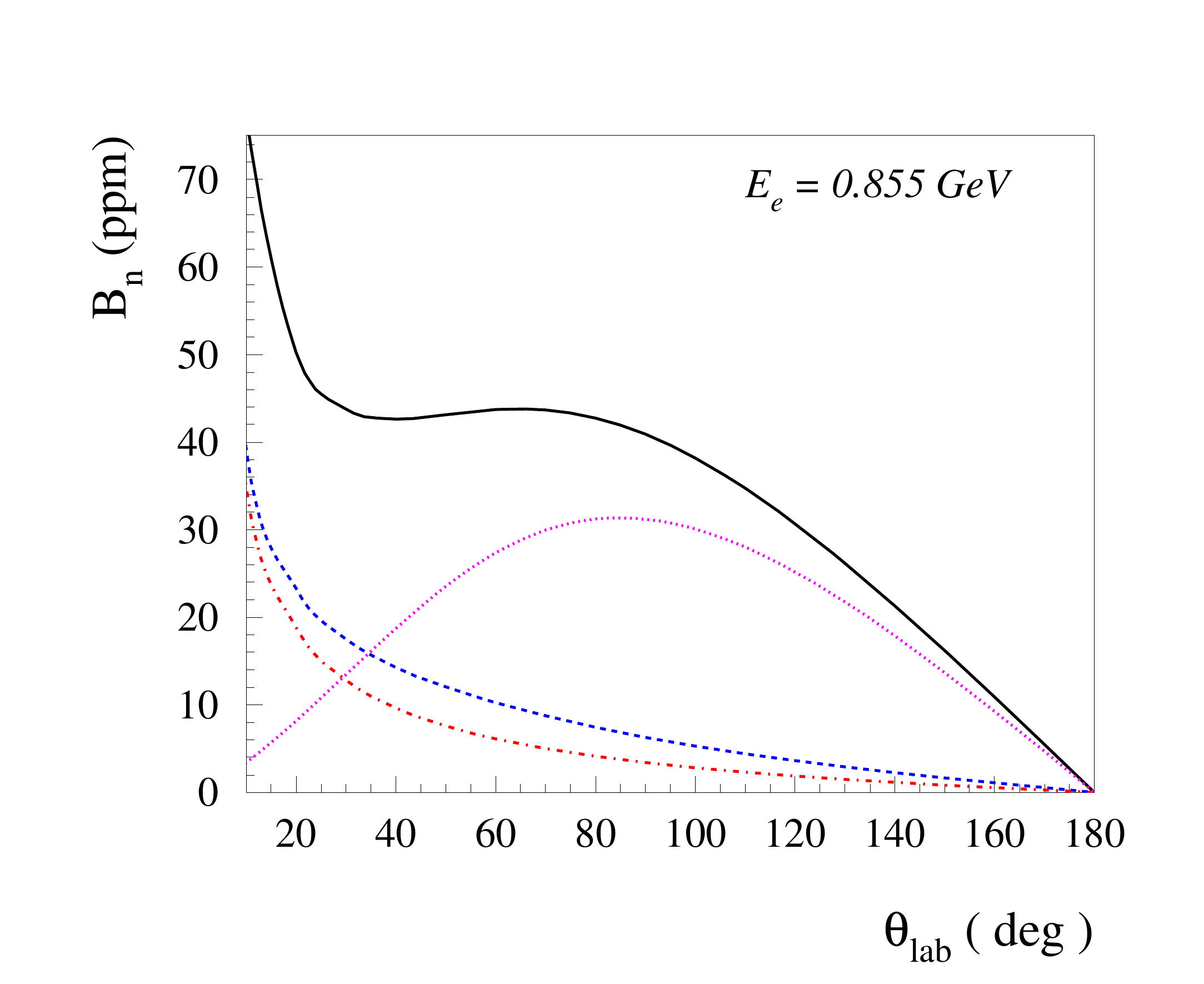}
\caption{Beam normal spin asymmetry $B_n$ for the $e^- p \to e^- \Delta^+$ process 
as function of the {\it lab} scattering angle for a beam energy $E_e = 0.855$~GeV 
where data have been taken by the A4@MAMI experiment~\cite{Maas:2004pd, Rios:2017vsw}.  
The curves denote the contributions from different intermediate states: 
nucleon (dashed-dotted red curve); 
$\Delta(1232)$ (dashed blue curve);
$S_{11}(1535) + D_{13}(1520)$ (dotted violet curve); 
$N + \Delta + S_{11}(1535) + D_{13}(1520)$ (solid black curve). 
}
\label{fig:a4mami1}
\end{figure}

Fig.~\ref{fig:a4mami2} shows the results for $B_n$ for two beam energies of the A4@MAMI experiment below the thresholds for $S_{11}(1535)$ and $D_{13}(1520)$. 
These kinematical situations are therefore dominated by $N$ and $\Delta$ intermediate state contributions. 
We see that the corresponding asymmetries become large at forward angles. In the angular range $\theta_{\rm lab} = 30 - 40$~deg, where potential data exist from the A4@MAMI experiment, we predict 
$B_n \simeq 200 - 250$~ppm for $E_e = 0.420$~GeV,  and 
$B_n \simeq 75 - 95$~ppm for $E_e = 0.570$~GeV.
It will be interesting to confront these numbers with experiment.

\begin{figure}
\includegraphics[width=9.cm]{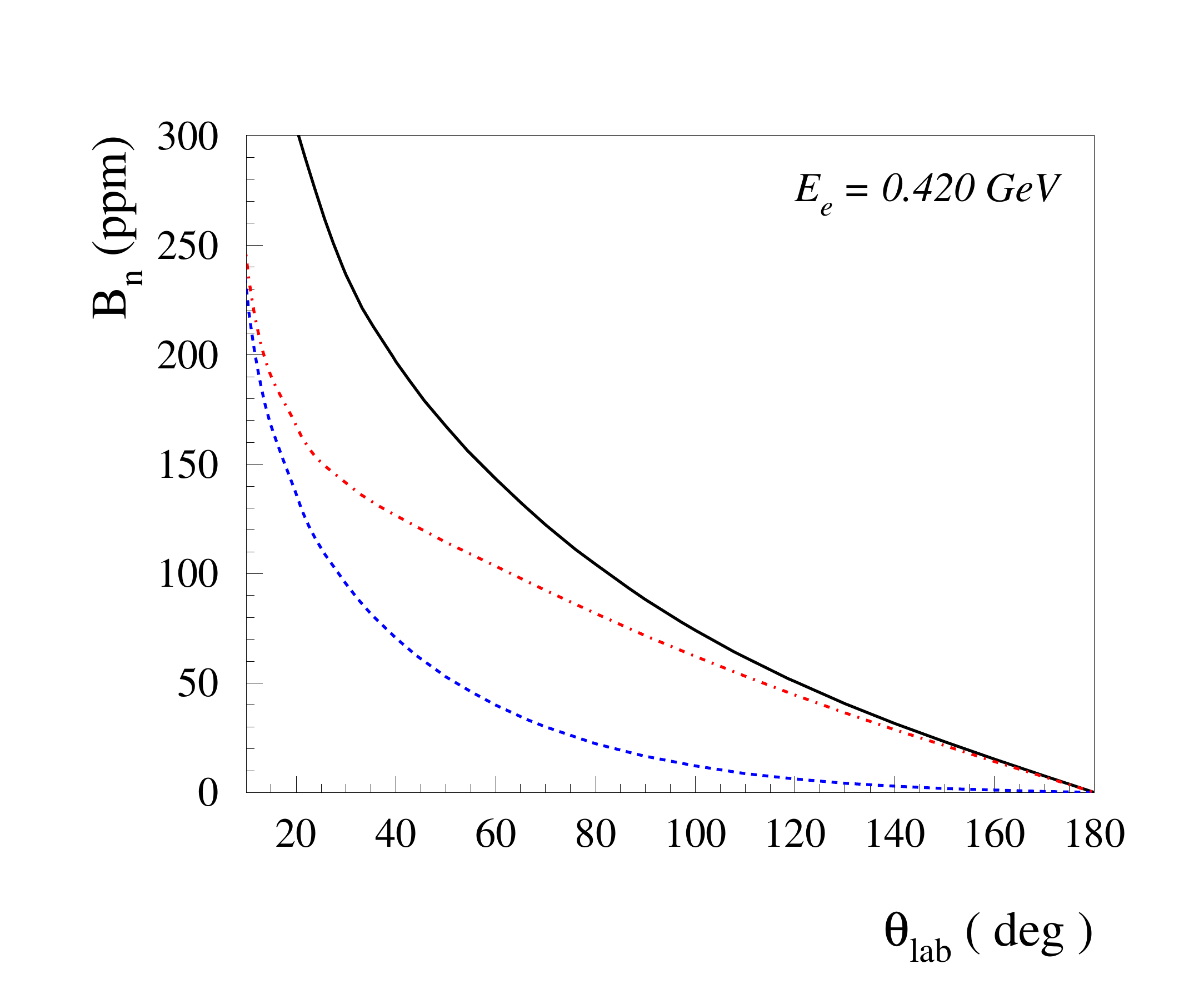}
\includegraphics[width=9.cm]{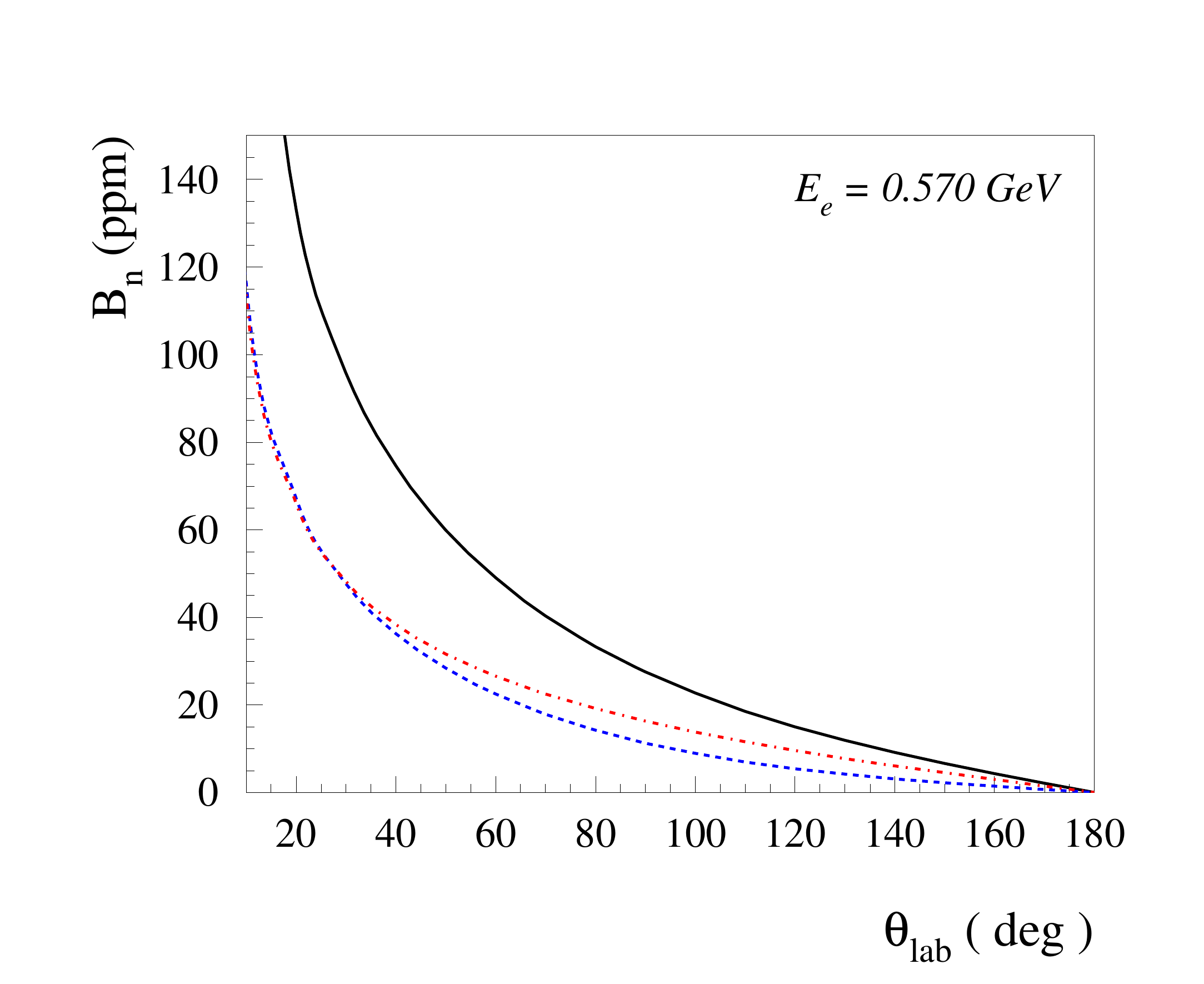}
\caption{Beam normal spin asymmetry $B_n$ for the $e^- p \to e^- \Delta^+$ process 
as function of the {\it lab} scattering angle, 
for beam energies in the $\Delta$-resonance region where data have been taken 
by the A4@MAMI experiment~\cite{Maas:2004pd, Rios:2017vsw}.   
Upper panel: $E_e = 0.420$~GeV; lower panel: $E_e = 0.570$~GeV). 
The curves denote the contributions from different intermediate states: 
nucleon (dashed-dotted red curves); 
$\Delta(1232)$ (dashed blue curves); 
$N + \Delta$ (solid black curves). 
}
\label{fig:a4mami2}
\end{figure}

In Fig.~\ref{fig:bn_sensitivity}, we also show the sensitivity of $B_n$ at $E_e = 0.570$~GeV to the value of the $\Delta^+$ magnetic dipole moment $\mu_\Delta$. We compare our results for three values of $\mu_\Delta$ corresponding with the theoretical uncertainty range which is currently listed by PDG, given in Eq.~(\ref{eq:mdm}). We see from Fig.~\ref{fig:bn_sensitivity} that for $\theta_{\rm lab}$ around 90$^\circ$, $B_n$ varies by around 5~ppm when varying $\mu_\Delta$ in the range $\mu_\Delta = 1.5 - 4.5$ (in units $e/(2 M_\Delta)$),  in a region where $B_n$ is about 28 ppm.
 
\begin{figure}
\includegraphics[width=9.cm]{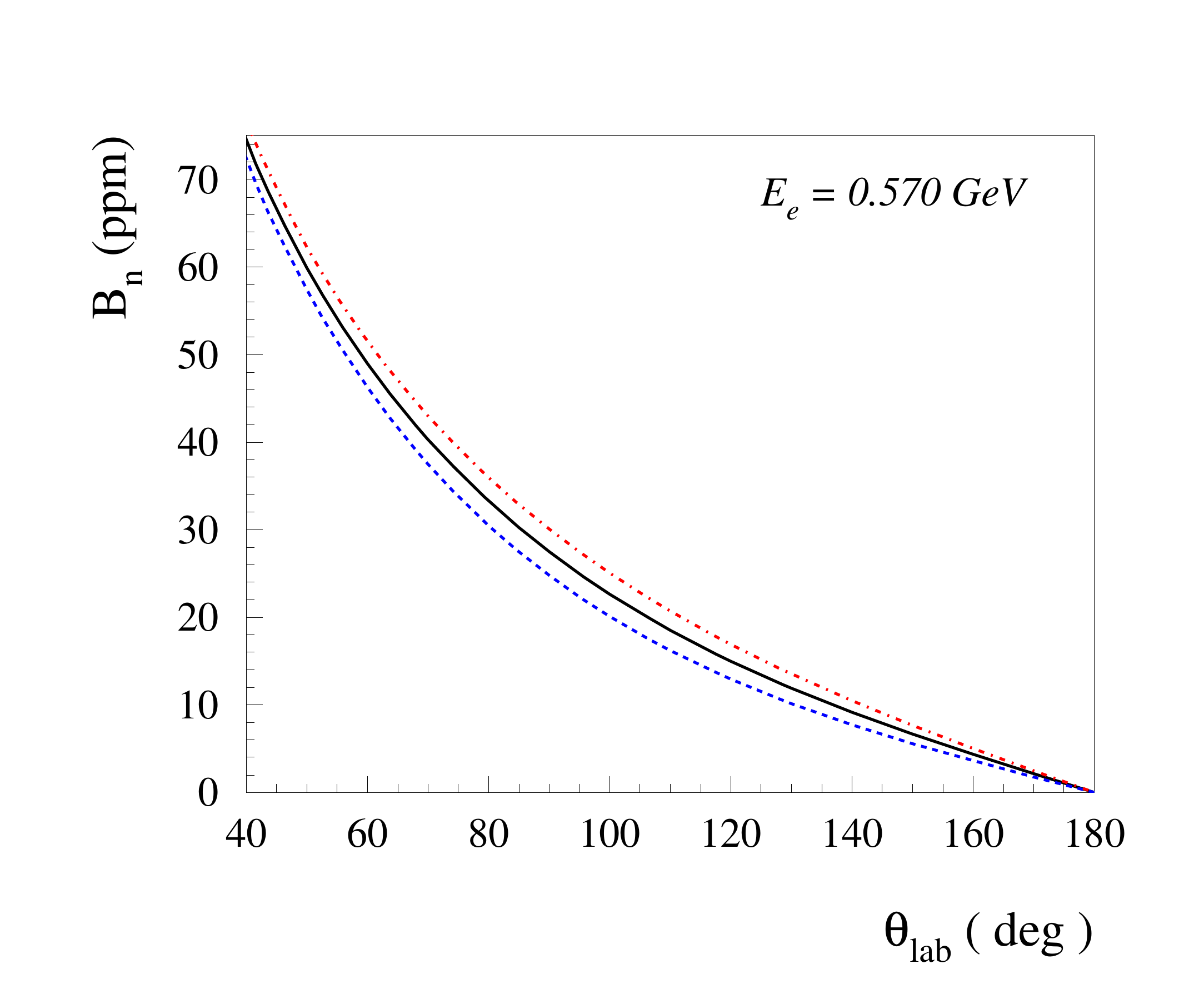}
\caption{Sensitivity of the beam normal spin asymmetry $B_n$ for the $e^- p \to e^- \Delta^+$ process 
at $E_e = 0.570$~GeV on the $\Delta^+$ magnetic dipole moment.  
The curves denote the contributions from $N + \Delta$ intermediate states 
for different values of  $\mu_\Delta$ (in units of $e/(2 M_\Delta)$):
$\mu_\Delta = 1.5$  (blue dashed curve); 
$\mu_\Delta = 3.0$  (black solid curve); 
$\mu_\Delta = 4.5$  (red dashed-dotted curve).
}
\label{fig:bn_sensitivity}
\end{figure}

\section{Conclusions}
\label{sec:concl}

In this work, we have presented the general formalism to describe the beam normal spin asymmetry $B_n$ for the $e p \to e \Delta^+(1232)$ process. This beam normal SSA arises from an interference between a one-photon exchange amplitude and the absorptive part of a two-photon exchange amplitude. As the intermediate state in the TPE amplitude is on its mass shell, it allows access to the $\Delta \to \Delta$ and $N^\ast \to \Delta$ electromagnetic transitions, which otherwise are not accessible in an experiment without resorting to a theory framework. We have provided estimates for this asymmetry by considering nucleon, $\Delta(1232)$, $S_{11}(1535)$, and 
$D_{13}(1520)$ intermediate states. We find that $B_n$ for the $e p \to e \Delta$ process shows a strong enhancement in the forward angular range, as compared to its counterpart for the elastic process $ep \to e p$, which has been measured by several collaborations. 
The forward enhancement of $B_n$ for the inelastic process is due to the OPE process for the $e p \to e \Delta$ process, entering the denominator of $B_n$, which is suppressed by one power of $Q^2$ relative to its elastic counterpart. The normal beam SSA for  the $e p \to e \Delta$ reaction therefore offers an increased sensitivity to the absorptive part of the TPE amplitude. 
We have compared our results for $B_n$ with the first data point for the $e^- p \to e^- \Delta^+$ process from the Qweak@JLab experiment and found that the forward angle data point 
is very well described both in sign and magnitude by our calculation. 
We have also given predictions for the A4@MAMI experiment, for which data have been taken, and have shown the sensitivity of this observable to the $\Delta^+$ magnetic dipole moment.   It will be interesting to analyse those data and provide a comparison with the  above theory predictions.

\begin{acknowledgments}
The authors are grateful to Lothar Tiator and Vladimir Pascalutsa for useful discussions. CEC thanks the National Science Foundation (USA) for support under grant PHY-1516509.  The work of MV is supported by the Deutsche Forschungsgemeinschaft DFG through the Collaborative Research Center CRC1044.  MV thanks the College of William and Mary for its hospitality during intermediate stages of this work and CC thanks the Johannes Gutenberg-University for hospitality during the completion of this work.
\end{acknowledgments}

\appendix*

\section{Electromagnetic $\Delta \to S_{11}$ and $\Delta \to D_{13}$ transitions in the quark model}

For calculations with the $S_{11}$ and $D_{13}$ intermediate states, we need the $\Delta \to S_{11}$ and $\Delta \to D_{13}$ transition matrix elements, as well as the proton to $S_{11}$ and $D_{13}$ matrix elements.  The latter are known from analyses of scattering with proton targets~\cite{Tiator:2011pw}, but for the former no direct experimental information is available.  

However, using ideas from $SU(6)$ or from the constituent quark model one can relate the transition matrix elements involving $\Delta$s to those involving nucleons.  We shall implement these ideas in a nonrelativistic (NR) limit, and give the helicity amplitudes for the transitions connecting a $\Delta$ to the $S_{11}$ or $D_{13}$ in terms of those connecting a proton to the same states.  A summary of the techniques and the relevant results are given here.  Details regarding the techniques can be found in~\cite{Close:1979bt}, and 
the same methods of course can be used for other transitions as well~\cite{Close:1979bt,Carlson:1998jk,Rislow:2010vi,Carlson:2012pc}.

The helicity matrix elements, defined for the present cases in Eqs.~\eqref{eq:SDhelicity} and~\eqref{eq:DDhelicity}, contain the operators $J_\mu \cdot \epsilon^\mu_{\lambda=1}$ and $J^0$.  At the quark level in a NR limit, these operators become
\begin{align}
J_\mu \cdot \epsilon^\mu_{\lambda=1} &\to 3 A e_{q3} S_{3+} + 3 B e_{q3} L_{3+}	\,,  \nn\\
J^0 &\to 3 C e_{q3}		\,.
\end{align}
The operators are written in anticipation of use in a wave function completely antisymmetric among the quarks, so we only evaluate for the third quark and multiply by 3; $e_{q3}$ is the charge of the third quark, $S_{3+}$ is the spin raising operator for the third quark, and 
$L_{3+}$ similarly is the angular momentum raising operator. We have let the photon three-momentum be in the $z$-direction.  The factors $A$, $B$, and $C$ depend on position; $C$ is the simplest example being just $e^{iq z_3}$ where $z_3$ is the $z$ coordinate of the third quark.  Details of the derivations may be found in~\cite{Close:1979bt} starting from a Hamiltonian formalism, and one can obtain the same results using a NR reduction of standard relativistic expressions for the current.  

The $\Delta$ state has the same spatial wave function as the nucleon state, and may in short form be given as
\begin{align}
\ket{ \Delta(S_z)} = \ket{ \psi^S_{00} \phi^S \chi^S_{S_z} } ,
\end{align}
where $\psi$, $\phi$, and $\chi$ respectively represent the space, flavor, and spin wave functions of the three quarks, the color wave function is tacit, superscripts $S$ indicate a wave function that is totally symmetric,  the subscripts on the space wave function indicate orbital angular momentum and projection, $L$ and $L_z$, and the subscript on the spin wave function is the spin projection.  The flavor wave function, here and elsewhere in this section, is chosen to be for the total charge $+1$ state.

The states $S_{11}(1535)$ and $D_{13}(1520)$ are negative parity states usually associated with the $SU(6)$ $70$-plet states where the three quarks are collectively in a spin-1/2, flavor octet state.  Mixing with other states is possible but will be ignored for now.  The wave functions, again in short form, are
\begin{align}
\ket{J,J_z} &= \frac{1}{2}	\sum_{L_z,S_z}
	\left(		\begin{array}{c|cc}
			J	&	1	&	1/2	\\
			J_z	&	L_z	&	S_z
			\end{array}
	\right)							\nn\\
&	\times
\Big\{		\psi^{MS}_{1 L_z}
		\left( - \phi^{MS} \chi^{MS}_{S_z}  + \phi^{MA} \chi^{MA}_{S_z}  \right)	\nn\\
&	\hskip 1 em	+	
		\psi^{MA}_{1 L_z}
		\left(  \phi^{MA} \chi^{MS}_{S_z}  + \phi^{MS} \chi^{MA}_{S_z}  \right)		
\Big\}	\,,
\end{align}
where $J$ is a stand-in for $S_{11}$ when $J=1/2$ or $D_{13}$ when $J=3/2$.  The first symbol after the summation sign is  Clebsch-Gordan coefficient, and superscripts $MS$ and $MA$ stand for mixed symmetry states where the first pair of quarks is either symmetric or antisymmetric.

The crucial matrix elements involving the spatial wave function of the ground state $N$ or $\Delta$ on one side and the mixed symmetry states of the $70$-plet on the other side are
\begin{align}
A_1(Q^2) &= \braket{ \psi^{MS}_{10} | A | \psi^S_{00}	}	,	\nn\\
B_1(Q^2) &= \braket{ \psi^{MS}_{11} | B L_{3+}  | \psi^S_{00}	}	,	\nn\\
C_1(Q^2) &= \braket{ \psi^{MS}_{10} | C | \psi^S_{00}	}	,
\end{align}
where $A_1$, $B_1$, and $C_1$ are generally real.  The $MA$ states do not enter because of  symmetry considerations.  Then,
\begin{align}
A^{\Delta S}_{1/2} &= \frac{ N_{S\Delta} }{ 3 \sqrt{3} } A_1(Q^2)	,	\nn\\
A^{\Delta S}_{-1/2} &= - \frac{ N_{S\Delta} }{ 3  } A_1(Q^2)	,	\nn\\
A^{\Delta D}_{3/2} &= 0	,	\nn\\
A^{\Delta D}_{1/2} &= - \frac{ N_{D\Delta} }{ 3  } \sqrt{ \frac{2}{3} }	A_1(Q^2)	,	\nn\\
A^{\Delta D}_{-1/2} &= - \frac{ N_{D\Delta} \sqrt{2}  }{ 3  } 	A_1(Q^2)	.
\label{eq:qmrel1}
\end{align}
The $B$ amplitudes also do not enter, because of the mismatched spins of the $\Delta$ and $S_{11}$, $D_{13}$ quark wave functions, meaning that the $S_{3+}$ operator is always needed. Similarly, all the scalar $S_{11}$ and $D_{13}$ transition amplitudes to the $\Delta$ are zero.  Normalizations $N_{S\Delta}$ and $N_{D\Delta}$ are given in Eqs.~\eqref{eq:NSDel} and~\eqref{eq:NDDel}, respectively. 

A pair of proton to $70$-plet amplitudes 
are 
\begin{align}
A_{1/2}^{pS} &= \frac{N_{pS}}{\sqrt{6}} \left( - A_1(Q^2) + \sqrt{2} B_1(Q^2) \right),	\nn\\
A_{1/2}^{pD} &= \frac{N_{pD}}{\sqrt{6}} \left( \sqrt{2} A_1(Q^2) + B_1(Q^2) \right).
\end{align}
These allow us to obtain $A_1$ from measured amplitudes,
\begin{align}
A_1(Q^2) = \sqrt{ \frac{2}{3} } \left( \frac { \sqrt{2} A^{pD}_{1/2} }{ N_{pD} }
		- \frac{ A^{pS}_{1/2} }{ N_{pS} }	\right).
\label{eq:qmrel2}
\end{align}

The MAID parameterizations are~\cite{Tiator:2011pw}:
\begin{align}
\label{eq:helmaid}
A^{pS}_{1/2} & = 66.4 \times 10^{-3} \, \text{GeV}^{-1/2}
	\left( 1 + 1.608 \, Q^2 \right) e^{-0.70 Q^2}	,	\nn\\
S^{pS}_{1/2} & = -2.0 \times 10^{-3} \, \text{GeV}^{-1/2}
	\left( 1 + 23.9 \, Q^2 \right) e^{-0.81 Q^2}	,	\nn\\
A^{pD}_{1/2} &= -27.4  \times 10^{-3} \, \text{GeV}^{-1/2}		\nn\\
	&\times \left( 1 +8.580 \,Q^2 - 0.252 \,Q^4 +0.357 \,Q^8 \right) e^{-1.20 Q^2},	\nn\\
A^{pD}_{3/2} &= 160.6  \times 10^{-3} \, \text{GeV}^{-1/2}		\nn\\
	&\times \left( 1 - 0.820 \,Q^2 + 0.541 \,Q^4 - 0.016 \,Q^8 \right) e^{-1.06 Q^2},	\nn\\
S^{pD}_{1/2} &= -63.5  \times 10^{-3} \, \text{GeV}^{-1/2}		
	 \left( 1 + 4.19 \,Q^2 \right) e^{-3.40 Q^2},	\nn \\
\end{align}
for $Q^2$ in GeV$^2$.


\begin{thebibliography}{99}

\bibitem{Pascalutsa:2006up} 
  V.~Pascalutsa, M.~Vanderhaeghen and S.~N.~Yang,
  Phys.\ Rept.\  {\bf 437}, 125 (2007). 

\bibitem{Alexandrou:2012da} 
  C.~Alexandrou, C.~N.~Papanicolas and M.~Vanderhaeghen,
  Rev.\ Mod.\ Phys.\  {\bf 84}, 1231 (2012).

\bibitem{Aznauryan:2012ba} 
  I.~G.~Aznauryan {\it et al.},
  Int.\ J.\ Mod.\ Phys.\ E {\bf 22}, 1330015 (2013). 

\bibitem{Jenkins:1994md} 
  E.~E.~Jenkins and A.~V.~Manohar,
  Phys.\ Lett.\ B {\bf 335}, 452 (1994).
    
\bibitem{Lebed:2004fj} 
  R.~F.~Lebed and D.~R.~Martin,
  Phys.\ Rev.\ D {\bf 70}, 016008 (2004). 
    
\bibitem{Buchmann:2000wf} 
  A.~J.~Buchmann and R.~F.~Lebed,
  Phys.\ Rev.\ D {\bf 62}, 096005 (2000).
      
\bibitem{Buchmann:2002et} 
  A.~J.~Buchmann and R.~F.~Lebed,
  Phys.\ Rev.\ D {\bf 67}, 016002 (2003).


\bibitem{Gellas:1998wx} 
  G.~C.~Gellas, T.~R.~Hemmert, C.~N.~Ktorides and G.~I.~Poulis,
  Phys.\ Rev.\ D {\bf 60}, 054022 (1999). 

\bibitem{Pascalutsa:2004je} 
  V.~Pascalutsa and M.~Vanderhaeghen,
  Phys.\ Rev.\ Lett.\  {\bf 94}, 102003 (2005). 

\bibitem{Gail:2005gz} 
  T.~A.~Gail and T.~R.~Hemmert,
  Eur.\ Phys.\ J.\ A {\bf 28}, 91 (2006).
  
\bibitem{Procura:2008ze} 
  M.~Procura,
  Phys.\ Rev.\ D {\bf 78}, 094021 (2008)
  
\bibitem{Ledwig:2011cx} 
  T.~Ledwig, J.~Martin-Camalich, V.~Pascalutsa and M.~Vanderhaeghen,
  Phys.\ Rev.\ D {\bf 85}, 034013 (2012).
    
    
\bibitem{Alexandrou:2008bn} 
  C.~Alexandrou {\it et al.},
  Phys.\ Rev.\ D {\bf 79}, 014507 (2009). 
  
\bibitem{Alexandrou:2009hs} 
  C.~Alexandrou, T.~Korzec, G.~Koutsou, C.~Lorce, J.~W.~Negele, V.~Pascalutsa, A.~Tsapalis and M.~Vanderhaeghen,
  Nucl.\ Phys.\ A {\bf 825}, 115 (2009). 

\bibitem{Aubin:2008qp} 
  C.~Aubin, K.~Orginos, V.~Pascalutsa and M.~Vanderhaeghen,
  Phys.\ Rev.\ D {\bf 79}, 051502 (2009).
  
  
\bibitem{Machavariani:1999fr}
  A.~I.~Machavariani, A.~Faessler and A.~J.~Buchmann,
  Nucl.\ Phys.\ A {\bf 646}, 231 (1999)
  [Erratum-ibid.\ A {\bf 686}, 601 (2001)].

\bibitem{Drechsel:2000um}
  D.~Drechsel, M.~Vanderhaeghen, M.~M.~Giannini and E.~Santopinto,
  Phys.\ Lett.\ B {\bf 484}, 236 (2000).

\bibitem{Drechsel:2001qu}
  D.~Drechsel and M.~Vanderhaeghen,
  Phys.\ Rev.\ C {\bf 64}, 065202 (2001).


\bibitem{Chiang:2004pw}
  W.~T.~Chiang, M.~Vanderhaeghen, S.~N.~Yang and D.~Drechsel,
  Phys.\ Rev.\ C {\bf 71}, 015204 (2005).


\bibitem{Pascalutsa:2007wb} 
  V.~Pascalutsa and M.~Vanderhaeghen,
  Phys.\ Rev.\ D {\bf 77}, 014027 (2008).
  
 \bibitem{Kotulla:2002cg} 
  M.~Kotulla {\it et al.},
  Phys.\ Rev.\ Lett.\  {\bf 89}, 272001 (2002). 
  
 \bibitem{Olive:2016xmw} 
  C.~Patrignani {\it et al.} [Particle Data Group],
  Chin.\ Phys.\ C {\bf 40}, no. 10, 100001 (2016).
  
\bibitem{Schumann:2010js} 
  S.~Schumann {\it et al.},
  Eur.\ Phys.\ J.\ A {\bf 43}, 269 (2010). 
  
  
      
\bibitem{Carlson:2007sp} 
  C.~E.~Carlson and M.~Vanderhaeghen,
  Ann.\ Rev.\ Nucl.\ Part.\ Sci.\  {\bf 57}, 171 (2007). 
  
  
\bibitem{DeRujula:1972te} 
  A.~De Rujula, J.~M.~Kaplan and E.~De Rafael,
  Nucl.\ Phys.\ B {\bf 35}, 365 (1971).

\bibitem{Zhang:2015kna} 
  Y.~W.~Zhang {\it et al.},
  Phys.\ Rev.\ Lett.\  {\bf 115}, no. 17, 172502 (2015).
  
  
\bibitem{Afanasev:2002gr} 
  A.~Afanasev, I.~Akushevich and N.~P.~Merenkov,
  hep-ph/0208260.
  
  
\bibitem{Gorchtein:2004ac} 
  M.~Gorchtein, P.~A.~M.~Guichon and M.~Vanderhaeghen,
  Nucl.\ Phys.\ A {\bf 741}, 234 (2004).

\bibitem{Pasquini:2004pv} 
  B.~Pasquini and M.~Vanderhaeghen,
  Phys.\ Rev.\ C {\bf 70}, 045206 (2004).
    
\bibitem{Kumar:2013yoa} 
  K.~S.~Kumar, S.~Mantry, W.~J.~Marciano and P.~A.~Souder,
  Ann.\ Rev.\ Nucl.\ Part.\ Sci.\  {\bf 63}, 237 (2013). 
  
\bibitem{Wells:2000rx} 
  S.~P.~Wells {\it et al.} [SAMPLE Collaboration],
  Phys.\ Rev.\ C {\bf 63}, 064001 (2001). 


\bibitem{Maas:2004pd} 
  F.~E.~Maas {\it et al.},
  Phys.\ Rev.\ Lett.\  {\bf 94}, 082001 (2005).

  \bibitem{Rios:2017vsw} 
  D.~B.~Ríos {\it et al.},
  Phys.\ Rev.\ Lett.\  {\bf 119}, no. 1, 012501 (2017).
  
  
\bibitem{Armstrong:2007vm} 
  D.~S.~Armstrong {\it et al.} [G0 Collaboration],
  Phys.\ Rev.\ Lett.\  {\bf 99}, 092301 (2007). 
  
\bibitem{Androic:2011rh} 
  D.~Androic {\it et al.} [G0 Collaboration],
  Phys.\ Rev.\ Lett.\  {\bf 107}, 022501 (2011). 
  
  
\bibitem{Abrahamyan:2012cg} 
  S.~Abrahamyan {\it et al.} [HAPPEX and PREX Collaborations],
  Phys.\ Rev.\ Lett.\  {\bf 109}, 192501 (2012).

\bibitem{Waidyawansa:2013yva} 
  B.~P.~Waidyawansa [Qweak Collaboration],
  AIP Conf.\ Proc.\  {\bf 1560}, 583 (2013).
    
\bibitem{Nuruzzaman:2015vba} 
  Nuruzzaman [Qweak Collaboration],
  arXiv:1510.00449 [nucl-ex].

\bibitem{Dalton:2015lqa} 
  M.~M.~Dalton,
  arXiv:1510.01582 [nucl-ex].

\bibitem{Buncher:2016nmv} 
  B.~Buncher and C.~E.~Carlson,
  Phys.\ Rev.\ D {\bf 93}, no. 7, 074032 (2016). 

\bibitem{Jones:1972ky} 
  H.~F.~Jones and M.~D.~Scadron,
  Annals Phys.\  {\bf 81}, 1 (1973).
  
 
  
\bibitem{Drechsel:2007if} 
  D.~Drechsel, S.~S.~Kamalov and L.~Tiator,
  Eur.\ Phys.\ J.\ A {\bf 34}, 69 (2007).

\bibitem{Tiator:2011pw} 
  L.~Tiator, D.~Drechsel, S.~S.~Kamalov and M.~Vanderhaeghen,
  Eur.\ Phys.\ J.\ ST {\bf 198}, 141 (2011). 

\bibitem{Guichon:1998xv} 
  P.~A.~M.~Guichon and M.~Vanderhaeghen,
  Prog.\ Part.\ Nucl.\ Phys.\  {\bf 41}, 125 (1998). 
  
 \bibitem{Weber:1978dh}
  H.~J.~Weber and H.~Arenhovel,
  Phys.\ Rept.\  {\bf 36}, 277 (1978).

\bibitem{Nozawa:1990gt} 
  S.~Nozawa and D.~B.~Leinweber,
  Phys.\ Rev.\ D {\bf 42}, 3567 (1990).

  
\bibitem{Pascalutsa:1998pw} 
  V.~Pascalutsa,
  Phys.\ Rev.\ D {\bf 58}, 096002 (1998).
  
\bibitem{Pascalutsa:1999zz} 
  V.~Pascalutsa and R.~Timmermans,
  Phys.\ Rev.\ C {\bf 60}, 042201 (1999).
  
 
\bibitem{Close:1979bt} 
  F.~E.~Close,
  ``An Introduction to Quarks and Partons,''
  Academic Press/London 1979, 481p


 \bibitem{Carlson:1998jk} 
  C.~E.~Carlson and C.~D.~Carone,
  Phys.\ Lett.\ B {\bf 441}, 363 (1998). 
   
\bibitem{Rislow:2010vi} 
  B.~C.~Rislow and C.~E.~Carlson,
  Phys.\ Rev.\ D {\bf 83}, 113007 (2011).
  
\bibitem{Carlson:2012pc} 
  C.~E.~Carlson and B.~C.~Rislow,
  Phys.\ Rev.\ D {\bf 86}, 035013 (2012).


\end{thebibliography}
\end{document}